\documentclass[aps,prd,nofootinbib,showpacs,eqsecnum,twocolumn,floatfix]{revtex4}
\usepackage{amssymb,amsfonts}
\usepackage{bm} % bold math
\usepackage{graphicx}
\newcommand{\dert}[2]{{\partial #1 / \partial #2}}

\newcommand{\w}[1]{\bm{#1}}
\newcommand{\be}{\begin{equation}}
\newcommand{\ee}{\end{equation}}
\newcommand{\bea}{\begin{eqnarray}}
\newcommand{\eea}{\end{eqnarray}}
\newcommand{\Hor}{{\mathcal H}}
\newcommand{\M}{{\mathcal M}}
\newcommand{\Sp}{{\mathcal S}}
\newcommand{\T}{{\mathcal T}}
\newcommand{\el}{\w{\ell}}

\newcommand{\dd}{\bm{\mathrm{d}}}
\newcommand{\Lie}[1]{\bm{\mathcal L}_{\w{#1}}\,}
\newcommand{\Liec}[1]{{\mathcal L}_{\w{#1}}\,}
\newcommand{\LieS}[1]{{}^{\Sp}\!\Lie{#1}}
\newcommand{\hajicek}{H\'a\'\j i\v{c}ek}

\newcommand{\equalH}{\stackrel{\Hor}{=}}
\newcommand{\vqs}{\vec{\w{q}}^*}
\newcommand{\DS}{\w{\mathcal{D}}}
\newcommand{\DSc}{\mathcal{D}}
\newcommand{\volS}{{}^{\scriptscriptstyle\mathcal{S}}\!\w{\epsilon}}
\newcommand{\volH}{{}^{\scriptscriptstyle\mathcal{H}}\!\w{\epsilon}}
\begin{document}

\title{A generalized Damour-Navier-Stokes equation applied to trapping 
horizons}

\author{Eric Gourgoulhon}
\email[]{eric.gourgoulhon@obspm.fr} 
\affiliation{Laboratoire de
l'Univers et de ses Th\'eories, UMR 8102 du C.N.R.S., Observatoire
de Paris, F-92195 Meudon Cedex, France}

\date{17 October 2005}

\begin{abstract}
An identity is derived from Einstein equation for any hypersurface 
$\Hor$ which can be
foliated by spacelike  two-dimensional surfaces.  In the case where the
hypersurface is null, this identity coincides with the two-dimensional
Navier-Stokes-like equation obtained by Damour in the membrane approach to a
black hole event horizon.  In the case where $\Hor$ is spacelike or null and 
the 2-surfaces are marginally trapped, this identity applies to  Hayward's
trapping horizons and to the related dynamical horizons recently introduced by
Ashtekar and Krishnan. The identity involves a normal fundamental form 
(normal connection 1-form) of the
2-surface, which can be viewed as a generalization to non-null hypersurfaces of
the \hajicek\  1-form used by Damour.  This 1-form is also used to define the
angular momentum of the horizon.  The generalized  Damour-Navier-Stokes
equation leads then to a simple evolution equation for the angular momentum.  
\end{abstract}


\maketitle

%%%%%%%%%%%%%%%%%%%%%%
\section{Introduction} \label{s:intro}

The concept of black hole shear viscosity has been introduced by 
Hawking and Hartle \cite{HawkiH72,Hartl73,Hartl74} when studying 
the response of the event horizon to external perturbations.
It was then greatly enhanced by Damour \cite{Damou78,Damou79,Damou82}
who showed that a 2-dimensional spacelike section of the event horizon
can be considered as a fluid bubble endowed with some mechanical 
and electromagnetic properties. Moreover, he 
explicitly derived from Einstein equation a Navier-Stokes equation,
involving both shear and bulk viscosities,  
for an effective momentum density of the ``fluid'' constituting the
bubble. This ``fluid bubble'' point of view lead to the development
of the so-called \emph{membrane paradigm} for black hole astrophysics
\cite{PriceT86,ThornPM86}. An example of recent work using the concept of 
black hole viscosity is the study of tidal interaction in 
binary-black-hole inspiral \cite{PriceW01}. 

The membrane approach was related to the event horizon of the black hole
(or to the associated ``stretched horizon'' \cite{PriceT86,ThornPM86}). 
However the event horizon is an extremely global and teleological concept,
which requires the knowledge of the complete spacetime, 
including the full future of any Cauchy surface, to be located.
It can by no means be determined from local measurements
(see \S~2.2.2 of Ref.~\cite{AshteK04} for an interesting example
of some event horizon which appears in a \emph{flat} region of spacetime).
This makes the event horizon a not very practical representation
of black holes for studies beyond the stationary regime in 
numerical relativity or quantum gravity. 
For this reason, \emph{local} characterizations 
of black holes have been introduced in the last decade
(see \cite{AshteK04,Booth05,GourgJ05} for a review). 
Although some local concepts, like Hawking's \emph{apparent horizon}
\cite{HawkiE73}, have appeared well before, the local approach has 
really started with Hayward's introduction of \emph{trapping horizons}
(or more precisely \emph{future outer trapping horizons})
\cite{Haywa94b}. Whereas apparent horizons are 2-dimensional surfaces 
(associated with some spacelike slicing of spacetime), 
trapping horizons are 3-dimensional
submanifolds of spacetime (hypersurfaces), as event horizons. 
Basically a trapping
horizon is a world tube made of marginally trapped 2-surfaces;
Hayward has studied the dynamics of these objects on their own,
without making any reference to any slicing of spacetime by spacelike
Cauchy surfaces. More recently, Ashtekar 
and Krishnan \cite{AshteK02,AshteK03,AshteK04}
have introduced the related concept of \emph{dynamical
horizons} and have established the ``first law'' of black hole
thermodynamics for them. This first law has been extended to 
trapping horizons \cite{Haywa04c,Haywa04b}. 

A natural question which arises is then: can the ``fluid bubble'' approach
to event horizons be extended to these local characterizations of
black holes ? In particular, can one obtain an analog of Damour's
Navier-Stokes equation~?
Although some viscosity aspects are already present (in form of 
dissipations terms) in the first law of dynamical horizons
established by Ashtekar, Krishnan and Hayward, it does not seem
a priori obvious to get an equivalent of the Navier-Stokes equation. 
In particular, Damour's derivation relied heavily on the null
structure of the event horizon, whereas the future
outer trapping horizons are generically spacelike in dynamical situations,
being null only in stationary states. 

We will show here that 
it is indeed possible to get a Navier-Stokes-like equation,
provided one introduces the correct geometrical objects.
Actually, the Navier-Stokes-like equation derived hereafter is quite general:
it applies not only to trapping or dynamical horizons, but to any
hypersurface which can be foliated by a smooth 
family of spacelike 2-surfaces. 
In this respect, the recent demonstration \cite{AshteG05} of the uniqueness
of the foliation of a given dynamical horizon by marginally trapped surfaces
is providing some motivation 
for the present work. Moreover, some uniqueness theorems about dynamical
and trapping horizons have been recently established \cite{AshteG05,AnderMS05},
conferring to these objects a more solid physical status. In addition,
the recent study \cite{BoothBGV05} has provided some deep insights about
their behavior in various scenarios of
gravitational collapse. 

The plan of the article is as follows. In Sec.~\ref{s:foliat}, we set
the basic framework of our study, namely a hypersurface
foliated by spacelike 2-surfaces. In Sec.~\ref{s:Extrinc_Spt} we review
standard results about the extrinsic geometry of a single spacelike 2-surface.
At this stage we introduce the geometrical
object that shall play the role of a momentum density in the Navier-Stokes
equation, namely a normal fundamental form of the 2-surface. 
In Sec.~\ref{s:Extrinc_foliat} are introduced other geometrical objects, 
defined by the foliation as a whole and not by a single
2-surface. Then we have all the tools to derive the generalized
Damour-Navier-Stokes equation in Sec.~\ref{s:DNS}. An application to
a general law of angular momentum balance is given in Sec.~\ref{s:angu_mom}.
Finally Sec.~\ref{s:concl} contains the concluding remarks.

%%%%%%%%%%%%%%%%%%%%%%%%%%%%%%%%%%%%%%%%%%%%%%%%%%%%%%%%%%%%%%%%%%%%%%%%%%%%%%

\section{Foliation of a hypersurface by spacelike 2-surfaces}
\label{s:foliat}

\subsection{General set-up}

\begin{figure}
\includegraphics[width=0.3\textwidth]{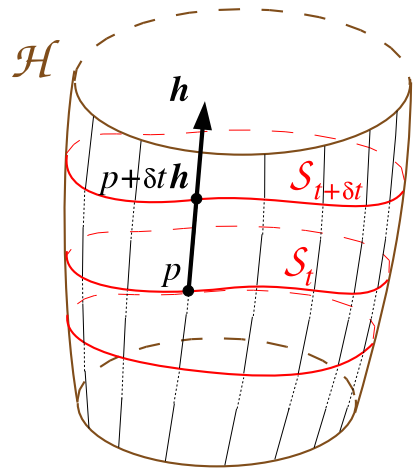} 
\caption{ \label{f:surf_foliat}
Foliation of a hypersurface $\Hor$ by a family $(\Sp_t)_{t\in\mathbb{R}}$
of spacelike 2-surfaces, and the associated evolution vector $\w{h}$.}
\end{figure}

We consider a spacetime $(\M,\w{g})$, i.e. a smooth manifold
$\M$ of dimension 4 endowed with a Lorentzian metric $\w{g}$, 
of signature $(-,+,+,+)$. We assume that $\M$ is time-orientable.
Let $\Hor$ be a hypersurface of $\M$ which is foliated by
a family $(\Sp_t)_{t\in\mathbb{R}}$ of 2-dimensional surfaces
$\Sp_t$ labeled by the real parameter $t$. 
By \emph{foliation}, it is meant that $\Hor = \bigcup_{t\in\mathbb{R}} \Sp_t$
and that for each point $p\in\Hor$, there is only one $\Sp_t$
going through $p$ (see Fig.~\ref{f:surf_foliat}). 
Then, given a coordinate system 
$x^a=(x^2,x^3)$ on each $\Sp_t$, $(t,x^2,x^3)$ constitutes a coordinate
system on $\Hor$. \footnote{Latin indices from the beginning of the alphabet
($a$, $b$, ...) run in $\{2,3\}$, Latin indices starting from the letter
$i$ run in $\{1,2,3\}$, whereas Greek indices run in $\{0,1,2,3\}$.} 
We assume that all surfaces $\Sp_t$ are \emph{spacelike} and \emph{closed}
(i.e. compact without boundary). 
In the framework of the 3+1 formalism of general relativity, 
one may think of each surface
$\Sp_t$ as being the intersection of $\Hor$ with a spacelike 
hypersurface $\Sigma_t$ arising from some 3+1 foliation 
$(\Sigma_t)_{t\in\mathbb{R}}$ of $\M$:
$\Sp_t = \Hor \cap \Sigma_t$. Such a viewpoint will be called hereafter
\emph{a 3+1 perspective} (e.g. \cite{GourgJ05}). \label{s:perspective}
It will not be used 
in the mainstream of this article, except for 
making remarks and connections with previous works. 
Indeed, we will deal 
only with quantities intrinsic to $\Hor$ and its
foliation $(\Sp_t)_{t\in\mathbb{R}}$.
Besides, let us recall that for a dynamical horizon the foliation 
$(\Sp_t)_{t\in\mathbb{R}}$ by marginally trapped surfaces is unique
(up to a relabeling $t\mapsto t'$) \cite{AshteG05}. 

Demanding that the 2-surface $\Sp_t$ is spacelike amounts to
saying that the metric $\w{q}$ induced by the spacetime metric $\w{g}$
onto $\Sp_t$ is positive definite (i.e. Riemannian). 
In particular $\w{q}$ is not degenerate and at each point $p\in\Sp_t$, 
the following orthogonal decomposition holds:
\be \label{e:TM_decomp}
    \T_p(\M) = \T_p(\Sp_t) \oplus \T_p(\Sp_t)^\perp , 
\ee
where $\T_p(\M)$ [resp. $\T_p(\Sp_t)$] denotes the space of vectors
tangent to $\M$ [resp. to $\Sp_t$] at the point $p$, and 
$\T_p(\Sp_t)^\perp$ denotes the space of vectors orthogonal
to $\Sp_t$ at $p$. Both vector spaces $\T_p(\Sp_t)$ and
$\T_p(\Sp_t)^\perp$ are two-dimensional. Let us then denote by $\w{\vec{q}}$ 
the orthogonal projector onto $\T_p(\Sp_t)$:
$\w{\vec{q}}(\w{v}) = \w{v} \iff \w{v}\in\T_p(\Sp_t)$ and
$\w{\vec{q}}(\w{v}) = 0 \iff \w{v}\in\T_p(\Sp_t)^\perp$.
In this article, we shall take a 4-dimensional point of view on 
the induced metric $\w{q}$ by setting
$\w{q}(\w{u},\w{v})=0$ if any of the two vectors $\w{u}$ and
$\w{v}$ in $\T_p(\M)$ belongs to $\T_p(\Sp_t)^\perp$. 
Then, if in a given basis, the components of $\w{q}$ are $q_{\alpha\beta}$, 
the components of $\vec{\w{q}}$ are $q^\alpha_{\ \, \beta}$, where
the index $\alpha$ has been raised with the metric $\w{g}$. 

Given a generic tensor $\w{A}$ on $\T_p(\M)$ of covariance type $(m,n)$,
we define a new tensor $\vqs\w{A}$ of the same covariance type
thanks to the projector $\w{\vec{q}}$:
\bea
   \left( \vec{q}^* A \right)^{\alpha_1\ldots\alpha_m}_{\qquad \ \  \beta_1
   \ldots\beta_n}
   & := & q^{\alpha_1}_{\ \ \mu_1} \ldots q^{\alpha_m}_{\ \ \mu_m} 
   q^{\nu_1}_{\ \ \beta_1} \ldots q^{\nu_n}_{\ \ \beta_n} \times
    \nonumber \\
 & & A^{\mu_1\ldots\mu_m}_{\qquad \ \   \nu_1\ldots\nu_n} .
\eea
Note that for a vector $\vqs\w{v} = \w{\vec{q}}(\w{v})$ and for
a 1-form, $\vqs\w{\omega} = \w{\omega} \circ \w{\vec{q}}$.
Note also that for multilinear forms intrinsic to the 2-dimensional 
manifold $\Sp_t$, $\vqs$ can be viewed as the ``push-forward''
operator which transforms them to multilinear forms acting on 
the 4-dimensional space $\T_p(\M)$ (for vectors and more generally
contravariant tensors, the push-forward 
operator is canonically provided by the embedding of $\Sp_t$ in $\M$). 
A tensor $\w{A}$ on $\M$ will be said \emph{tangent
to $\Sp_t$} if $\vqs\w{A} = \w{A}$.

\subsection{Evolution vector} \label{s:evol_vect}

Let us denote by $\w{h}$ the vector field on $\Hor$ 
such that (see Fig.~\ref{f:surf_foliat})
(i) $\w{h}$ is tangent to $\Hor$, (ii) at any point in $\Hor$, 
$\w{h}$ is orthogonal to the surface $\Sp_t$ going through this point, 
(iii) the length of $\w{h}$ is associated with the parameter $t$ labeling
the surfaces $(\Sp_t)$ by
\be \label{e:Lieh_t_1}
    \Lie{h} t = 1 , 
\ee
where $\Lie{h}$ denotes the Lie derivative along $\w{h}$. In the
present case (scalar field $t$), we have of course
$\Lie{h} t  = h^\mu \partial_\mu t = \langle \dd t, \w{h}\rangle$, 
where brackets are used to denote the action of 1-forms on vectors. 
Given the foliation $(\Sp_t)_{t\in\mathbb{R}}$, 
the conditions (i), (ii) and (iii) define $\w{h}$ uniquely.
Note however, that if the leaves $\Sp_t$ are relabeled by a new parameter
$t'= F(t)$ (where $F:\, \mathbb{R}\rightarrow\mathbb{R}$ is a smooth
one-to-one map), then $\w{h}$ is transformed into
\be
    \w{h'} = [F'(t)]^{-1} \, \w{h}  
\ee 
so that $\Lie{h'}t'=1$. 

An immediate consequence of Eq.~(\ref{e:Lieh_t_1}) is that the 2-surfaces
$\Sp_t$ are \emph{Lie-dragged} by the vector field $\w{h}$: given an
infinitesimal parameter $\delta t$, 
the image of the surface  $\Sp_t$ by the displacement
of each of its points by the vector $\delta t \, \w{h}$ 
is the surface $\Sp_{t+\delta t}$
(cf. Fig.~\ref{f:surf_foliat}). For this reason, $\w{h}$ is the natural vector field
to describe the ``evolution'' of quantities across the foliation 
of $\Hor$. In particular, we will consider the Lie derivative along $\w{h}$,
$\Lie{h}$ as the ``evolution operator''\footnote{The term
``evolution'' stands for ``variation as $t$ increases'' and can 
be made more concrete is one adopts the 3+1 perspective mentioned in 
Sec.~\ref{s:perspective}.} along $\Hor$. Since $\w{h}$ Lie-drags the
surfaces $(\Sp_t)$, it transports any vector tangent to $\Sp_t$
to a vector tangent to $\Sp_{t+\delta t}$. In other words,
\be \label{e:Lieh_v_Sp}
    \forall \w{v}\in\T(\Sp_t),\  \Lie{h} \w{v} \in \T(\Sp_t) ,
\ee
where $\T(\Sp_t)$ denotes the space of vector fields defined
on $\Hor$ and which are tangent to $\Sp_t$. 
Although the vector field $\w{h}$ is not tangent to $\Sp_t$, 
we can use property (\ref{e:Lieh_v_Sp}) to extend
the definition of $\Lie{h}$ to 1-forms $\w{\omega}$ acting in $\T(\Sp_t)$
(i.e. 2-dimensional 1-forms associated with the manifold structure of
$\Sp_t$), by setting
\be 
  \forall \w{v}\in\T(\Sp_t),\quad 
  \langle \Lie{h}\w{\omega}, \w{v} \rangle :=
  \Lie{h} \langle \w{\omega}, \w{v} \rangle 
  - \langle \w{\omega}, \Lie{h} \w{v} \rangle .
\ee
Note that the right-hand side of this equation is well defined thanks
to Eq.~(\ref{e:Lieh_v_Sp}). The definition of $\Lie{h}$ is then 
extended immediately to any tensor field on $\Sp_t$ via tensor products
and Leibniz' rule, e.g. $\Lie{h} (\w{\omega}_1 \otimes \w{\omega}_2)
:= \Lie{h}\w{\omega}_1\otimes \w{\omega}_2
+ \w{\omega}_1\otimes \Lie{h} \w{\omega}_2$.
Given a multilinear form field $\w{A}$ on $\Sp_t$, we then denote
by $\LieS{h}\w{A}$ the push-forward (via the projector $\w{\vec{q}}$)
of the derivative $\Lie{h}\w{A}$ defined above:
\be
    \LieS{h}\w{A} := \vqs \Lie{h}\w{A} . 
\ee
One can then show that (see Appendix~A of Ref.~\cite{GourgJ05} for
details):
\be \label{e:LieS_Lie}
    \LieS{h}\w{A} = \vqs \Lie{h} \vqs \w{A} , 
\ee
where the Lie derivative in the right-hand side is the standard 
Lie derivative along $\w{h}$ within the manifold $\M$. 
 
Owing to the fundamental property (\ref{e:Lieh_t_1})
and the resulting Lie-dragging of the surfaces $(\Sp_t)$, 
it is not surprising that the vector $\w{h}$ has been introduced by
many authors when studying foliation of hypersurfaces, in various
contexts: $\w{h}$ and $\Lie{h}$ were denoted respectively $\el$ and
$D/dt$ by Damour \cite{Damou78,Damou79,Damou82} in his 
black hole mechanics ($\Hor$ was then taken to be an event horizon);
$\w{h}$ was denoted $\partial^{(n)}_t$ by Eardley \cite{Eardl98} in
his study of black hole boundary conditions for 3+1 numerical relativity, 
since it was then viewed as the part of the evolution vector $\dert{}{t}$ 
which is normal to $\Sp_t$ in a coordinate system adapted to $\Hor$. 
Similarly, $\w{h}$ is denoted $\vec{\zeta}$ by Cook \cite{Cook02} 
when searching for boundary conditions for initial data representing
quasi-stationary black holes. More recently, in the context
of trapping and dynamical horizons, $\w{h}$ has been denoted
$\overline{V}^a$ by Ashtekar and Krishnan \cite{AshteK03},
$\w{\mathcal{V}}$ by Booth and Fairhurst \cite{BoothF04,BoothF05}
and $\w{\xi}$ by Hayward \cite{Haywa04c,Haywa04b}.

Let $C$ be the scalar field defined on $\Hor$ as half 
the scalar square of $\w{h}$:
\be \label{e:def_C}
    C := \frac{1}{2} \, \w{h} \cdot \w{h} , 
\ee
where a dot is used to denote the scalar product taken with the metric
$\w{g}$. Since $\w{h}$ is normal to $\Sp_t$, an orthogonal vector basis of 
$\T_p(\Hor)$ is $(\w{h},\w{e}_2,\w{e}_3)$, where $(\w{e}_2,\w{e}_3)$ is
an orthonormal basis of $\T_p(\Sp_t)$. In this basis, the matrix of the 
metric induced by $\w{g}$ on $\Hor$ is 
$\mathrm{diag}(2C,1,1)$. We then conclude that 
\be
    \begin{array}{lclcl}
    \Hor\ \mbox{is spacelike} & \iff & C > 0 & \iff & \w{h}\  \mbox{is spacelike}\\ 
    \Hor\ \mbox{is null} & \iff & C = 0 & \iff & \w{h}\  \mbox{is null}\\ 
    \Hor\ \mbox{is timelike} & \iff & C < 0 & \iff & \w{h}\ \mbox{is timelike.}
    \end{array}
\ee

%%%%%%%%%%%%%%%%%%%%%%%%%%%%%%%%%%%%%%%%%%%%%%%%%%%%%%%%%%%%%%%%%%%%%%%%%%%%%%%

\section{Extrinsic geometry of a spacelike 2-surface}
\label{s:Extrinc_Spt}

In this section, we review some basic results about the extrinsic geometry
of a single spacelike 2-surface --- not necessarily a member of a
foliation. For future purpose, 
we take care to provide rather general definitions, for instance
not limiting the definition of expansion and shear to null vectors,
as usually done,
nor limiting the definition of the normal fundamental forms to
some privileged normal frame.  

\subsection{Expansion and shear along normal vectors}

Let us consider a fixed 2-surface $\Sp_t$. We denote by $\T(\Sp_t)^\perp$ 
the space of vector fields $\w{v}$ which are defined on $\Sp_t$ and 
everywhere normal to $\Sp_t$:
$\forall p\in\Sp_t,\ \w{v}(p)\in \T_p(\Sp_t)^\perp$.
For any $\w{v}\in \T(\Sp_t)^\perp$,
we define the \emph{deformation tensor of $\Sp_t$ along $\w{v}$} as
the bilinear form
\be \label{e:def_deform}
    \w{\Theta}^{(\w{v})} := \vqs \w{\nabla} \underline{\w{v}} 
    \quad \left[\mbox{or}\ 
    \Theta^{(\w{v})}_{\alpha\beta} := \nabla_\nu v_\mu 
    \, q^\mu_{\ \, \alpha} q^\nu_{\ \, \beta} \right] , 
\ee
where $\w{\nabla}$ is the affine connection associated with the spacetime
metric $\w{g}$ and  
the underlining is used to denote in an index-free way the
1-form $\underline{\w{v}}$ canonically associated to the vector field
$\w{v}$ by the metric $\w{g}$. Note that thanks to the projector $\w{\vec{q}}$
in Eq.~(\ref{e:def_deform}), $\w{\Theta}^{(\w{v})}$ is independent of the
values of $\w{v}$ away from $\Sp_t$ (some extension of $\w{v}$ in an open
neighborhood of $\Sp_t$ being required for the spacetime covariant derivative
$\w{\nabla} \underline{\w{v}}$ to be well defined). 
It is easy to see that the bilinear form $\w{\Theta}^{(\w{v})}$ is symmetric, 
as the consequence of $\w{v}$ being normal to the surface $\Sp_t$ (Weingarten
property). 

Let us consider the metric $\w{\tilde q}$ induced by $\w{g}$ on 
the 2-surfaces deduced from $\Sp_t$ by Lie-dragging along  $\w{v}$
(recall that $\w{q}$ is a priori defined only on $\Sp_t$; we have of
course $\w{\tilde q} \stackrel{\Sp_t}{=} \w{q}$). Taking into account
the symmetry of $\w{\Theta}^{(\w{v})}$ and expressing the Lie derivative
in terms of $\w{\nabla}$ yields $
    \vqs \Lie{v} \w{\tilde q} = \vqs \w{\nabla}_{\w{v}} \w{\tilde q}
        + 2 \w{\Theta}^{(\w{v})} $.
Now, from the idempotent character of $\w{\vec{q}}$, it is easy
to see that $\vqs \w{\nabla}_{\w{v}} \w{\tilde q}=0$, so 
that finally one ends with
\be \label{e:Theta_v_Lie_q}
    \w{\Theta}^{(\w{v})} = \frac{1}{2} \, \vqs \Lie{v} \w{\tilde q} .
\ee
This equality justifies the name \emph{deformation tensor} given to 
$\w{\Theta}^{(\w{v})}$: $\w{\Theta}^{(\w{v})}$ measures the variation
of the metric in $\Sp_t$ when this surface is Lie-dragged along 
the vector $\w{v}$. 
Decomposing $\w{\Theta}^{(\w{v})}$ into a trace part and a traceless part
results in the definition of the \emph{expansion rate of 
$\Sp_t$ along $\w{v}$}:
\be \label{e:def_theta_v}
    \theta^{(\w{v})} := q^{\mu\nu} \Theta^{(\w{v})}_{\mu\nu}
    = \Lie{v} \ln \sqrt{\tilde q} , 
\ee
and the \emph{shear tensor of $\Sp_t$ along $\w{v}$}:
\be \label{e:def_shear}
    \w{\sigma}^{(\w{v})} := \w{\Theta}^{(\w{v})} - \frac{1}{2} \, 
    \theta^{(\w{v})} \, \w{q} . 
\ee
In Eq.~(\ref{e:def_theta_v}) the second equality results 
from Eq.~(\ref{e:Theta_v_Lie_q}), $\tilde q$ being the determinant of 
the components $\tilde q_{ab}$ with respect to
a coordinate system $x^a=(x^2,x^3)$ of the induced metric 
on the surface obtained from $\Sp_t$
by Lie-drag along $\w{v}$.    
Since $\sqrt{q}$ is related to the surface element 
$\volS$
of $\Sp_t$ by $\volS = \sqrt{q}\,  \dd x^2 \wedge \dd x^3$,
we see by considering coordinates $x^a$ constant along $\w{v}$ field
lines that $\theta^{(\w{v})}$ is nothing but the relative rate of change
of the area of a surface element Lie-dragged by $\w{v}$ from $\Sp_t$:
\be \label{e:Liev_volS}
    \Lie{\w{v}} \volS = \theta^{(\w{v})} \, \volS, 
\ee
hence the name \emph{expansion rate} given to $\theta^{(\w{v})}$. 

\begin{figure}
\includegraphics[width=0.45\textwidth]{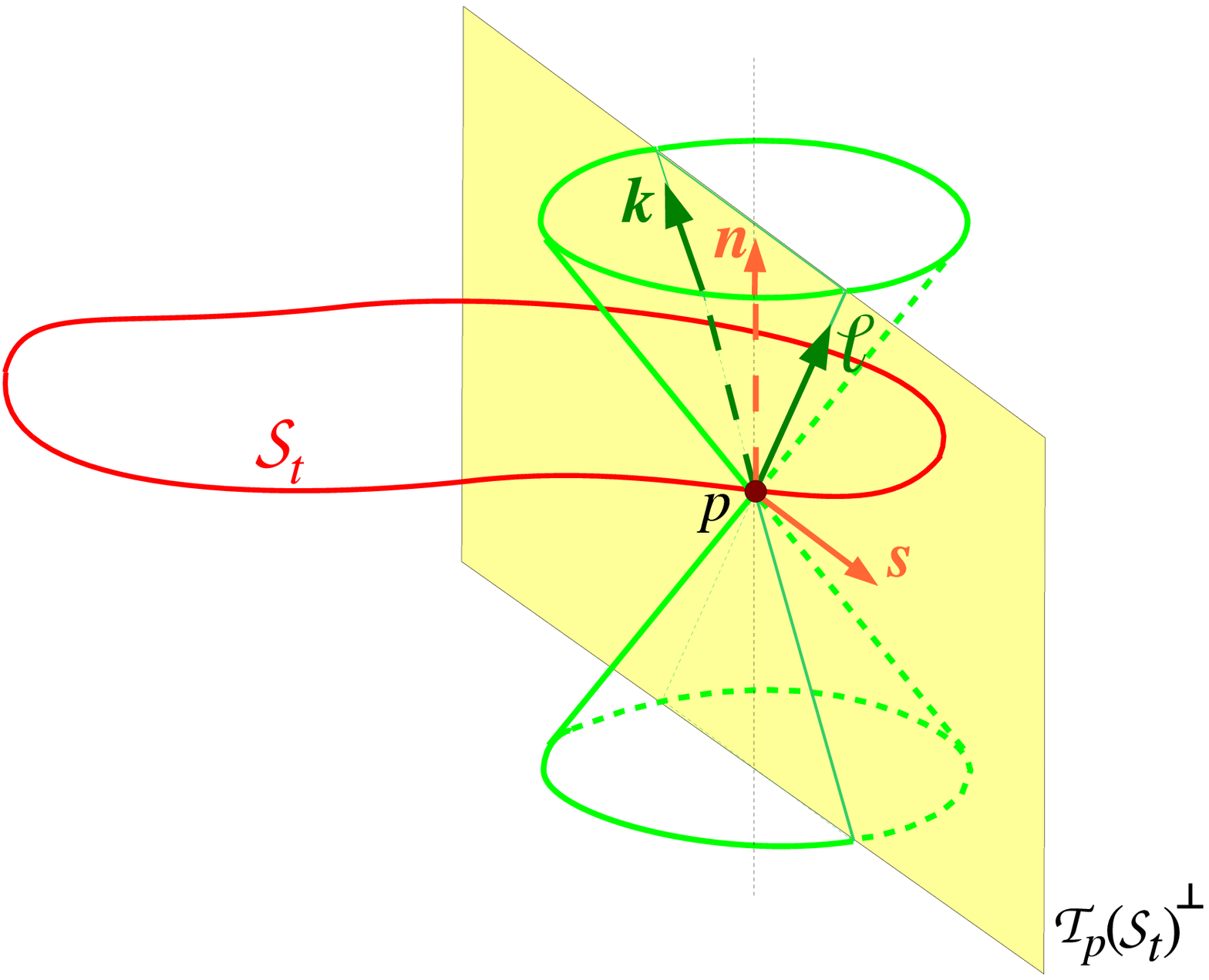} 
\caption{ \label{f:normal_plane}
Vector plane $\T_p(\Sp_t)^\perp$ normal to $\Sp_t$ at a given point $p$,
with some orthonormal frame $(\w{n},\w{s})$ and some
null frame $(\el,\w{k})$. The directions of $\el$ and $\w{k}$
are uniquely defined as the intersections of $\T_p(\Sp_t)^\perp$ with the
light cone emanating from $p$, whereas the directions of $\w{n}$ and $\w{s}$
can be changed by an arbitrary boost in a direction normal to $\Sp_t$.}
\end{figure}

\subsection{Normal frames} \label{s:normal_frame}

From the decomposition (\ref{e:TM_decomp}) and the spacelike character
of $\Sp_t$, we see that the restriction of the metric $\w{g}$ to the
vector plane $\T_p(\Sp_t)^\perp$ orthogonal to $\Sp_t$ must
be of signature $(-,+)$. There are then two natural choices of pairs
of vectors for generating this plane:
(i) an orthonormal basis $(\w{n},\w{s})$, i.e. a timelike vector $\w{n}$
and a spacelike vector $\w{s}$ satisfying
\be \label{e:scalar_ns}
    \w{n}\cdot\w{n} = -1,\quad \w{s}\cdot\w{s}=1,\quad
    \w{n}\cdot\w{s} = 0 ;
\ee
(ii) a pair of linearly independent future-directed
null vectors $(\el,\w{k})$; this choice is permissible since 
the signature $(-,+)$  implies that $\T_p(\Sp_t)^\perp$ contains 
two null directions, which  are actually the intersections of the 
null cone emanating from $p$
with $\T_p(\Sp_t)^\perp$ (cf. Fig.~\ref{f:normal_plane}):
\be \label{e:scalar_lk}
    \el\cdot\el = 0,\quad \w{k}\cdot\w{k}=0,\quad
    \el\cdot\w{k} =: - e^\sigma,
\ee 
where $\el\cdot\w{k}$ is negative (hence written
as minus some exponential) as a result
of both $\el$ and $\w{k}$ being future-directed. 

In both cases, there is not a unique choice: in case (i), the timelike
and spacelike directions can be changed by a boost in an arbitrary
direction normal to $\Sp_t$, leading to a new pair of basis vectors:
\bea
    \w{n'} & = & \cosh \eta \, \w{n} + \sinh \eta \, \w{s} \label{e:boost1} \\
    \w{s'} & = & \sinh \eta \, \w{n} + \cosh \eta \, \w{s} , \label{e:boost2}
\eea
where $\eta\in\mathbb{R}$ is the boost parameter. 
The choice $(\w{n},\w{s})$ can be made unique by invoking some 
extra structure like the global foliation $(\Sigma_t)_{t\in\mathbb{R}}$ 
of $\M$ arising from the
3+1 perspective mentioned in Sec.~\ref{s:perspective}, 
$\w{n}$ being then the future directed unit normal to $\Sigma_t$ and
$\w{s}$ one of the two unit normals to $\Sp_t$ which are 
tangent to $\Sigma_t$. Another definite choice of $(\w{n},\w{s})$ can
be performed when $\Hor$ is spacelike (resp. timelike), 
by demanding that $\w{n}$ (resp. $\w{s}$) is normal
to $\Hor$; 
$\w{s}$ (resp. $\w{n}$) is then colinear to the evolution vector $\w{h}$, 
and in particular lies in $\Hor$.  
This is the choice adopted by Ashtekar and Krishnan 
\cite{AshteK02,AshteK03,AshteK04} 
for dynamical horizons, which are always spacelike 
(see Appendix~\ref{s:AK}).   
 
In case (ii), the two null directions are unique, but the vectors
$\el$ and $\w{k}$ can be rescaled arbitrarily by
\bea
    \w{\ell'} & = & \lambda \, \el  , \qquad \lambda>0  \label{e:rescale_l}\\
    \w{k'} & = & \mu \, \w{k} , \qquad \mu>0 \label{e:rescale_k} ,
\eea
where the positive sign of $\lambda$ and $\mu$ is
chosen to preserve the future orientation. 
One may reduce the arbitrariness by fixing the
scalar product $\el\cdot\w{k}$ to $-1$ [choice $\sigma=0$
in Eq.~(\ref{e:scalar_lk})], but this determines only 
$\mu$ as being $\lambda^{-1}$ and leaves the degree of freedom on $\lambda$. 
We will see in Sec.~\ref{s:norm_frame_h} that, when considering
not a single $\Sp_t$, but the whole foliation $(\Sp_t)_{t\in\mathbb{R}}$,
this ambiguity can be fixed in a natural way, leading
to a unique choice of $(\el,\w{k})$.
 
Note that for the choice (i), the orthogonal projector $\w{\vec{q}}$
on $\Sp_t$ is expressible as
\be \label{e:q_ns}
    \w{\vec{q}} = \w{1} + \langle \underline{\w{n}}, . \rangle \, \w{n}
- \langle \underline{\w{s}}, . \rangle \, \w{s}
\ee
(equivalently 
$\w{q} = \w{g} + \underline{\w{n}} \otimes \underline{\w{n}}
        - \underline{\w{s}} \otimes \underline{\w{s}}$),
whereas for the choice (ii)
\be \label{e:q_kl}
\w{\vec{q}} = \w{1} + e^{-\sigma} \langle \underline{\w{k}}, . \rangle \, \el
+ e^{-\sigma} \langle \underline{\el}, . \rangle \, \w{k}
\ee
(equivalently 
$\w{q} = \w{g} + e^{-\sigma} \, \underline{\w{k}} \otimes \underline{\el}
    + e^{-\sigma} \, \underline{\el} \otimes \underline{\w{k}}$). 

$\Sp_t$ is called a \emph{trapped surface} if, in addition of being spacelike
and closed, it satisfies $\theta^{(\el)} <0$ and
$\theta^{(\w{k})} <0$, and a \emph{marginally trapped surface}
if $\theta^{(\el)} =0$ and
$\theta^{(\w{k})} <0$, or $\theta^{(\el)} <0$ and
$\theta^{(\w{k})}=0$ \cite{Penro65}. If one of the two null directions, $\el$ say,
can be selected as being ``outgoing'', $\Sp_t$ is called an
\emph{outer trapped surface} 
if $\theta^{(\el)} <0$
(irrespectively of the sign of $\theta^{(\w{k})}$) \cite{HawkiE73}. 
It is called a
\emph{marginally outer trapped surface (MOTS)} if $\theta^{(\el)}=0$.
Notice that all these definitions are unaffected by the rescaling
(\ref{e:rescale_l})-(\ref{e:rescale_k}) of the null vectors $\el$ and
$\w{k}$. 

\subsection{Second fundamental tensor}

As for any non-null submanifold of $(\M,\w{g})$, the \emph{second fundamental
tensor} of $\Sp_t$ (also called \emph{extrinsic imbedding curvature tensor}
\cite{Carte92a} or \emph{shape tensor} \cite{Senov04}) is defined
as the tensor $\w{\mathcal{K}}$ of type $(1,2)$ relating the covariant derivative of 
a vector tangent to $\Sp_t$ taken with the spacetime connection $\w{\nabla}$
to that taken with the connection in $\Sp_t$ compatible with the induced
metric $\w{q}$, hereafter denoted by $\DS$:
\be
    \forall (\w{u},\w{v}) \in \T(\Sp_t)^2,\quad
    \w{\nabla}_{\w{u}} \w{v} = \DS_{\w{u}} \w{v} 
        + \w{\mathcal{K}}(\w{u},\w{v}). 
\ee
From the fundamental relation 
\be \label{e:DS_nabla}
    \DS \w{A} = \vqs \w{\nabla} \w{A} ,
\ee
valid for any tensorial field $\w{A}$ tangent to $\Sp_t$, it is easy 
to express $\w{\mathcal{K}}$ in terms of the derivative of $\w{\vec{q}}$:
\be \label{e:2FF_grad_q}
    \mathcal{K}^\alpha_{\ \, \beta\gamma} = 
        \nabla_\mu q^\alpha_{\ \, \nu} \; 
        q^\mu_{\ \, \beta} q^\nu_{\ \, \gamma} . 
\ee
Let $(\w{n},\w{s})$ be an orthonormal frame of $\T(\Sp_t)^\perp$;
inserting expression (\ref{e:q_ns}) for $q^\alpha_{\ \, \nu}$
in the above relation and making use of definition 
(\ref{e:def_deform}) leads to
\be \label{e:2FF_ns}
    \mathcal{K}^\alpha_{\ \, \beta\gamma} = 
    n^\alpha\, \Theta^{(\w{n})}_{\beta\gamma}
    - s^\alpha\, \Theta^{(\w{s})}_{\beta\gamma} .
\ee
Similarly, if one uses instead a null frame $(\el,\w{k})$
for $\T(\Sp_t)^\perp$, expression (\ref{e:q_kl}) 
for $q^\alpha_{\ \, \nu}$ leads to 
\be \label{e:2FF_kl}
    \mathcal{K}^\alpha_{\ \, \beta\gamma} = 
    e^{-\sigma} \left( k^\alpha\, \Theta^{(\el)}_{\beta\gamma}
    + \ell^\alpha\, \Theta^{(\w{k})}_{\beta\gamma} \right).   
\ee
It is clear on formulas (\ref{e:2FF_ns}) and (\ref{e:2FF_kl}) that
the second fundamental tensor is orthogonal to $\Sp_t$ in its
first index, and symmetric and tangent to $\Sp_t$ in its second and
third indices. 
The reader more familiar with the hypersurface case should note
that the second fundamental tensor for a hypersurface writes,
instead of Eq.~(\ref{e:2FF_ns}), 
$\mathcal{K}^\alpha_{\ \, \beta\gamma} = - n^\alpha K_{\beta\gamma}$,
where $n^\alpha$ is the normal to the hypersurface and $K_{\beta\gamma}$
its \emph{second fundamental form} or \emph{extrinsic curvature tensor}. 

\subsection{Normal fundamental forms}

Contrary to the case of a hypersurface, the extrinsic geometry of 
the 2-surface $\Sp_t$ is not entirely specified by the second
fundamental tensor $\w{\mathcal{K}}$. Indeed, because it involves
only the deformation tensors $\w{\Theta}^{(.)}$ of the normals to 
$\Sp_t$ [cf. Eqs.~(\ref{e:2FF_ns}) and (\ref{e:2FF_kl})],
$\w{\mathcal{K}}$ encodes only the part of the 
variation of $\Sp_t$'s normals which is parallel to $\Sp_t$. It 
does not encode the variation of the two normals with respect to each other. 
The latter is devoted to the \emph{normal fundamental forms}
which are the 1-forms defined by (cf. e.g. \cite{Haywa94b} or \cite{Epp00})
\bea
    \w{\Omega}^{(\w{n})} & := &\w{s} \cdot \w{\nabla}_{\w{\vec{q}}}\, \w{n} 
    \quad \left[\mbox{or}\ \Omega^{(\w{n})}_\alpha := s_\mu \nabla_\nu n^\mu \, 
        q^\nu_{\ \, \alpha} \right] 
            \label{e:def_Omega_n} \\
    \w{\Omega}^{(\w{s})} & := & \w{n} \cdot \w{\nabla}_{\w{\vec{q}}}\,  \w{s}
    \quad \left[\mbox{or}\ \Omega^{(\w{s})}_\alpha := n_\mu \nabla_\nu s^\mu \, 
        q^\nu_{\ \, \alpha}  \right]
\eea
if one considers an orthonormal frame $(\w{n},\w{s})$ of $\T(\Sp_t)^\perp$
and by
\be 
    \w{\Omega}^{(\el)} :=  \frac{1}{\w{k}\cdot\el} \, 
        \w{k} \cdot \w{\nabla}_{\w{\vec{q}}}\,  \el 
    \quad \left[\mbox{or}\ \Omega^{(\el)}_\alpha := \frac{1}{k_\rho \ell^\rho}
    k_\mu \nabla_\nu \ell^\mu \, 
        q^\nu_{\ \, \alpha} \right]   \label{e:def_Omega_l}
\ee
\be
    \w{\Omega}^{(\w{k})} :=  \frac{1}{\w{k}\cdot\el} \, 
        \el \cdot \w{\nabla}_{\w{\vec{q}}}\,  \w{k}  
    \quad \left[\mbox{or}\ \Omega^{(\w{k})}_\alpha := \frac{1}{k_\rho \ell^\rho}
    \ell_\mu \nabla_\nu k^\mu \, 
        q^\nu_{\ \, \alpha} \right]  \label{e:def_Omega_k}
\ee
if a null frame $(\el,\w{k})$ of $\T(\Sp_t)^\perp$ is considered instead. 
Note that, thanks to the projector $\w{\vec{q}}$, the definition of
the normal fundamental forms does not depend upon the values of the normal
fields $\w{n}$, $\w{s}$, $\el$ and $\w{k}$ away from the 2-surface $\Sp_t$. 
Note also that thanks to the division by $\w{k}\cdot\el$, the value
of $\w{\Omega}^{(\el)}$ does not depend on the choice of the null 
vector $\w{k}$ complementary to $\w{\el}$. 
From the orthogonality relations (\ref{e:scalar_ns}) and (\ref{e:scalar_lk}),
we have the immediate properties:
\bea
    \w{\Omega}^{(\w{s})} & = & - \w{\Omega}^{(\w{n})} \\
    \w{\Omega}^{(\w{k})} & = & - \w{\Omega}^{(\el)} + \DS \sigma . 
        \label{e:Omega_k_Omega_l}
\eea
We can also relate the $(\el,\w{k})$-type normal fundamental forms to the 
$(\w{n},\w{s})$-type ones by choosing the canonical null frame associated
with a given orthonormal frame $(\w{n},\w{s})$, namely
$\el = \w{n} + \w{s}$ and $\w{k} = \w{n} - \w{s}$. Then
\bea
    \w{\Omega}^{(\el)}  &=& \w{\Omega}^{(\w{n})} \qquad 
         [ \el = \w{n} + \w{s} ] \label{e:Omega_el_n} \\
    \w{\Omega}^{(\w{k})} &=&  - \w{\Omega}^{(\w{n})} \qquad 
        [ \w{k} = \w{n} - \w{s} ].
\eea
Note that, if one considers a non-null hypersurface
instead of a 2-surface, the analog of definition
(\ref{e:def_Omega_n}) would be 
$\w{\Omega}^{(\w{n})} :=\w{n} \cdot \w{\nabla}_{\w{\vec{q}}}\, \w{n}$,
since there is only one normal $\w{n}$. But this expression vanishes
identically by virtue of the normalization of $\w{n}$
($\w{n}\cdot\w{n}=1$ for a timelike hypersurface, and $-1$ for a spacelike
one). Consequently, the extrinsic curvature of a non-null hypersurface
is entirely characterized by the second fundamental form $\w{K}$.
For a null hypersurface, with normal $\el$, the orthogonal
projector $\w{\vec{q}}$ is not defined (as a result of $\w{q}$ being
degenerate). The relevant quantity is then 
the 1-form $\w{\Omega}^{(\el)}$ defined by Eq.~(\ref{e:def_Omega_l})
but with $\w{\vec{q}}$ substituted with the orthogonal projector to some
spacelike 2-surface embedded in the hypersurface and $\w{k}$ substituted 
with a transverse null vector. It is then called the \emph{\hajicek\  1-form}
\cite{Hajic73,Hajic75} (see also \cite{GourgJ05}).

The normal fundamental forms can be interpreted in terms of the
connection 1-forms associated with respect to some tetrad. Indeed
let $\w{e}_\alpha = (\w{n},\w{s},\w{e}_2,\w{e}_3)$ be an orthonormal
tetrad [$(\w{e}_2,\w{e}_3)$ is then an orthonormal basis of $\T_p(\Sp_t)$]. 
The \emph{connection 1-forms} associated with this tetrad are the
1-forms $\w{\omega}^\beta_{\ \, \alpha}$ such that for any vector field
$\w{v}$ on $\M$, $\w{\nabla}_{\w{v}} \, \w{e}_\alpha
        = \langle \w{\omega}^\mu_{\ \, \alpha}, \w{v}\rangle \, \w{e}_\mu $.
Then, from Eq.~(\ref{e:def_Omega_n}),
\be \label{e:Omega_connect_form}
    \w{\Omega}^{(\w{n})} = \vqs \w{\omega}^1_{\ \, 0} .
\ee
An equivalent phrasing of this is saying that the two non-trivial 
components of $\w{\Omega}^{(\w{n})}$ with respect to the dual frame
$(\w{e}^\alpha)$, namely $\Omega^{(\w{n})}_a$ ($a=2,3$),
 are identical to some of the connection coefficients
$\Gamma^\alpha_{\ \, \beta\gamma}$ associated with the tetrad 
$(\w{e}_\alpha)$:
\be \label{e:Omega_connect_coeff}
    \Omega^{(\w{n})}_a = \Gamma^1_{\ \, 0a} . 
\ee 
Relations (\ref{e:Omega_connect_form}) and (\ref{e:Omega_connect_coeff})
justify the alternative names \emph{external rotation coefficients} 
\cite{Carte92a}
and \emph{connection on the normal bundle} \cite{Epp00,BoothF05,Szaba04} 
given to the normal fundamental forms. 

The normal fundamental forms depend on the normal frame. 
Indeed a change of normal frame
$(\w{n},\w{s})\mapsto (\w{n'},\w{s'})$
according to Eqs.~(\ref{e:boost1})-(\ref{e:boost2}) leads to 
\be
    \w{\Omega}^{(\w{n'})} = \w{\Omega}^{(\w{n})} + \DS \eta , 
\ee
whereas a change of null normal frame 
$(\el,\w{k})\mapsto (\w{\ell'},\w{k'})$ according to
Eqs.~(\ref{e:rescale_l})-(\ref{e:rescale_k}) leads to 
\be \label{e:rescale_Omega_l}
    \w{\Omega}^{(\w{\ell'})} = \w{\Omega}^{(\el)} + \DS \ln \lambda .      
\ee
On the contrary, the second fundamental tensor $\w{\mathcal{K}}$ introduced
in the previous section does not depend on the choice of the normal frame:
this is obvious from Eq.~(\ref{e:2FF_grad_q}) which involves only
the projector $\w{\vec{q}}$, and this can be checked easily from 
the expressions in term of
the normal frames [Eqs.~(\ref{e:2FF_ns}) and (\ref{e:2FF_kl})],
by substituting the transformation laws (\ref{e:boost1})-(\ref{e:boost2}) 
and (\ref{e:rescale_l})-(\ref{e:rescale_k}).
We refer the reader to Carter's article 
\cite{Carte92a} for an extended discussion of this dependence of the
normal fundamental forms with respect to the normal frames.  

%%%%%%%%%%%%%%%%%%%%%%%%%%%%%%%%%%%%%%%%%%%%%%%%%%%%%%%%%%%%%%%%%%%%%%%%%%%%%%%

\section{Extrinsic geometry of the foliation}
\label{s:Extrinc_foliat}

Section~\ref{s:Extrinc_Spt} introduced quantities relative to a single
2-surface $\Sp_t$. Here we investigate quantities defined with respect
to the family $(\Sp_t)_{t\in\mathbb{R}}$ foliating $\Hor$. 

\subsection{Dual-null description of the foliation $(\Sp_t)_{t\in\mathbb{R}}$}

A very convenient way to study the foliation $(\Sp_t)_{t\in\mathbb{R}}$
is to employ of the dual-null formalism
of Hayward \cite{Haywa93,Haywa94b,Haywa01} (see also 
\cite{InverS80,BradyDIM96}), which we recall here, adapting the notations
to our purpose (see Table~\ref{t:Hayward} for the correspondence 
with Hayward's notations).

Let us consider, in the neighborhood of $\Hor$, two families of
null hypersurfaces, $(\mathcal{U}_u)_{u\in\mathbb{R}}$ and
$(\mathcal{V}_v)_{v\in\mathbb{R}}$, which intersect in spatial
2-surfaces such that each 2-surface $\Sp_t$ is one of these 
intersections, 
i.e. for each $t\in\mathbb{R}$, there exists a value of $u$, $u_0(t)$ say,
and a value of $v$, $v_0(t)$ say, such that $\Sp_t$ is the intersection
between $\mathcal{U}_{u_0(t)}$ and $\mathcal{V}_{v_0(t)}$: 
\be
    \forall t\in \mathbb{R},\quad \Sp_t = \mathcal{U}_{u_0(t)}
    \cap \mathcal{V}_{v_0(t)} .
\ee
Such a dual-null foliation always exists, $\mathcal{U}_{u_0(t)}$
(resp. $\mathcal{V}_{v_0(t)}$) being nothing but the hypersurface generated by
light rays outgoing (resp. ingoing) orthogonally from $\Sp_t$. 
Moreover, if $\Hor$ is spacelike or timelike, the dual-null foliation 
is unique. If $\Hor$ is null, then $\Hor$ coincides with 
$\mathcal{U}_{u_0(t)}$ and $u_0(t)=\mathrm{const}.$; there is then the
degree of freedom of choosing the foliation $(\mathcal{U}_u)$ outside
of $\Hor$. 

Let $\w{\tilde\ell}$ and $\w{\tilde k}$ be the null normal vectors 
to respectively $\mathcal{U}_u$ and $\mathcal{V}_v$ and dual (up to a sign)
to the gradient 1-forms $\dd u$ and $\dd v$:
\be \label{e:def_tilde_lk}
    \underline{\w{\tilde\ell}} := - \dd u \qquad \mbox{and}
    \qquad \underline{\w{\tilde k}} := - \dd v .
\ee
Since $\Sp_t$ belongs to both $\mathcal{U}_{u_0(t)}$
and $\mathcal{V}_{v_0(t)}$, both $\w{\tilde\ell}$ and $\w{\tilde k}$
are normal to $\Sp_t$, and therefore constitute a null frame
of $\T(\Sp_t)^\perp$, similar to those considered in Sec.~\ref{s:normal_frame}.
Let us denote $f$ the scalar field $\sigma$ associated to the
scalar product of $\w{\tilde\ell}$ and $\w{\tilde k}$ by 
Eq.~(\ref{e:scalar_lk}):
\be \label{e:tl_dot_tk}
    \w{\tilde\ell} \cdot \w{\tilde k} =: - e^{f} .  
\ee
From the definition (\ref{e:def_tilde_lk}), the 1-forms $\underline{\w{\tilde\ell}}$
and $\underline{\w{\tilde k}}$ are closed: 
\be \label{e:dtl_dtk}
    \dd \underline{\w{\tilde\ell}} = 0
    \qquad \mbox{and} \qquad
    \dd \underline{\w{\tilde k}} = 0.
\ee 
The vectors $\w{\tilde \ell}$ and $\w{\tilde k}$ being null, 
it follows immediately from Eq.~(\ref{e:dtl_dtk}) that
\be
    \w{\nabla}_{\w{\tilde \ell}} \, \w{\tilde \ell} = 0
    \qquad \mbox{and} \qquad
    \w{\nabla}_{\w{\tilde k}} \, \w{\tilde k} = 0 , 
\ee
i.e. the field lines of $\w{\tilde \ell}$ and $\w{\tilde k}$ are
geodesics: they are the light rays emanating orthogonally from $\Sp_t$. 
Hayward \cite{Haywa93,Haywa94b,Haywa01,Haywa04c,Haywa04b}
introduces another pair of null vectors by setting
\be \label{e:def_hat_lk}
    \w{\hat\ell} := e^{-f} \w{\tilde \ell}
    \qquad \mbox{and} \qquad
    \w{\hat k} := e^{-f} \w{\tilde k} .
\ee
These vectors have the fundamental property of Lie-dragging the 
hypersurfaces $\mathcal{V}_v$ and $\mathcal{U}_u$ respectively, i.e. they obey
\be  \label{e:lie_el_v}
    \Lie{\w{\hat\ell}} v = 1
    \qquad \mbox{and} \qquad
    \Lie{\w{\hat k}} u = 1 ,
\ee
but contrary to $\underline{\w{\tilde\ell}}$ and $\underline{\w{\tilde k}}$,
the 1-forms $\underline{\w{\hat\ell}}$ and $\underline{\w{\hat k}}$
are not closed, since we deduce from Eq.~(\ref{e:dtl_dtk}) that
\be
    \dd \underline{\w{\hat\ell}} = - \dd f \wedge 
     \underline{\w{\hat\ell}}
    \qquad \mbox{and} \qquad
    \dd \underline{\w{\hat k}} = - \dd f \wedge 
     \underline{\w{\hat k}} .
\ee 
Besides, 
\be
    \w{\hat\ell}\cdot\w{\hat k} = - e^{-f} . 
\ee

The \emph{anholonomicity 1-form}, also called \emph{twist 1-form},
is defined by [cf. Appendix~B of 
Ref.~\cite{Haywa94b}, in conjunction with our definitions
(\ref{e:def_Omega_l}) and (\ref{e:def_Omega_k})]:
\be
    \w{\varpi} := \frac{1}{2} \left( \w{\Omega}^{(\w{\tilde k})}
        - \w{\Omega}^{(\w{\tilde \ell})} \right) . 
\ee
From Eq.~(\ref{e:Omega_k_Omega_l}) with $\sigma=f$, we get
\be
    \w{\varpi} = - \w{\Omega}^{(\w{\tilde \ell})} + \frac{1}{2}
        \DS f = \w{\Omega}^{(\w{\tilde k})} - \frac{1}{2} \DS f . 
\ee
According to the scaling law (\ref{e:rescale_Omega_l}) with
$\lambda = e^{-f}$ [cf. Eq.~(\ref{e:def_hat_lk})], we can re-express
the anholonomicity 1-form in terms of the normal fundamental forms
associated with $\w{\hat\ell}$ and $\w{\hat k}$:
\be
    \w{\varpi} = - \w{\Omega}^{(\w{\hat \ell})} - \frac{1}{2} \DS f 
        = \w{\Omega}^{(\w{\hat k})} + \frac{1}{2} \DS f . 
\ee
Thanks to Eq.~(\ref{e:dtl_dtk}), we can easily re-express $\w{\varpi}$
in term of the commutator of $\w{\tilde k}$ and $\w{\tilde \ell}$,
as well as that of $\w{\hat k}$ and $\w{\hat \ell}$:
\be \label{e:anholon_commut}
    \w{\varpi} = \frac{e^{-f}}{2} \, 
    \w{q} \cdot [\w{\tilde k},\w{\tilde \ell}]
    = \frac{e^{f}}{2} \, \w{q} \cdot [\w{\hat k},\w{\hat \ell}] 
\ee
\be
    \left[ \mbox{or}\quad \varpi_\alpha = \frac{e^{-f}}{2} \, 
        q_{\alpha\mu} [\tilde k,\tilde \ell]^\mu
        = \frac{e^{f}}{2} \, q_{\alpha\mu} [\hat k,\hat \ell]^\mu
        \right] . 
\ee
This justifies the terms \emph{anholonomicity} and \emph{twist}
given to $\w{\varpi}$:
according to Frobenius theorem, the 2-planes $\T_p(\Sp_t)^\perp$ are integrable
in 2-surfaces when $t$ varies, if, and only if, the commutator of two 
generating vectors, e.g. $\w{\tilde k}$ and $\w{\tilde \ell}$, satisfies
$[\w{\tilde k},\w{\tilde \ell}] \in \T_p(\Sp_t)^\perp$; from
Eq.~(\ref{e:anholon_commut}), this is equivalent to $\w{\varpi}=0$. 

\begin{table}
\caption{\label{t:Hayward} Correspondence between our notations
and those of Hayward.}
\begin{ruledtabular}
\begin{tabular}{lll}
  this work & Hayward \cite{Haywa94b} &
  Hayward \cite{Haywa01,Haywa04c,Haywa04b} \\
\hline
$\w{q}$ & $h$ & $h$  \\
$\DS$ & $\mathcal{D}$ & $D$ \\
$u$ & $\xi_-$ & $x^-$ \\
$v$ & $\xi_+$ & $x^+$ \\
$\w{\tilde\ell}$ & $N_-$ & $g^{-1}(n^-)$ \\
$\w{\tilde k}$ & $N_+$ & $g^{-1}(n^+)$ \\
$\underline{\w{\tilde\ell}}$ & $n_-$ & $n^-$ \\
$\underline{\w{\tilde k}}$ & $n_+$ & $n^+$ \\
$f$ & $f$ & $f$ \\
$\w{\hat\ell}$ & $u_+-r_+ = e^{-f} N_-$ & $l_+$ \\
$\w{\hat k}$ & $u_- -r_- = e^{-f} N_+$ & $l_-$ \\
$\w{\varpi}$ & $\omega$ & $\omega$ \\
$\w{\Omega}^{(\w{\tilde \ell})}$ & $-e^{-f} \beta_-$ & $\zeta_{(+)}$ \\ 
$\w{\Omega}^{(\w{\tilde k})}$ & $-e^{-f} \beta_+$ & $\zeta_{(-)}$ \\ 
$\w{\Omega}^{(\w{\hat \ell})}$ & $e^{-f} \beta_+$ & $-\zeta_{(-)}$ \\ 
$\w{\Omega}^{(\w{\hat k})}$ & $e^{-f} \beta_-$ & $-\zeta_{(+)}$ \\ 
$\w{h}$ & & $\xi$ \\
$B$ & & $1/\xi^+$ \\
$C/A$ & & $-\xi^-$ \\
$\w{m}$ & & $\tau$ \\
\end{tabular}
\end{ruledtabular}
\end{table}

\subsection{Normal null frame associated with the evolution vector $\w{h}$}
\label{s:norm_frame_h}

\begin{figure}
\includegraphics[width=0.3\textwidth]{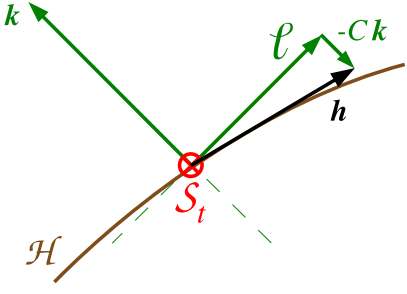} 
\caption{ \label{f:coupe_normal}
Null vectors $(\el,\w{k})$ associated with the evolution vector $\w{h}$
by $\w{h} = \el - C \w{k}$; 
the plane of the figure is the plane $\T_p(\Sp_t)^\perp$, so that
$\Sp_t$ is reduced to a point.}
\end{figure}

The evolution vector $\w{h}$ introduced in Sec.~\ref{s:evol_vect} 
belongs to the plane orthogonal to $\Sp_t$. 
Following Booth and Fairhurst \cite{BoothF04,BoothF05}, we notice that
there exists a unique pair of null vectors $(\el,\w{k})$
in that plane such that (see Fig.~\ref{f:coupe_normal})
\be \label{e:h_el_k}
    \w{h} = \el - C \w{k} 
    \qquad\mbox{and}\qquad
    \el\cdot\w{k} = - 1 ,
\ee
where $C$ is related to the scalar square of $\w{h}$ by 
Eq.~(\ref{e:def_C}). 
Thus we may say that the foliation
$(\Sp_t)_{t\in\mathbb{R}}$ entirely fixes, via its evolution
vector $\w{h}$, the ambiguities in the choice of the null normal
frame $(\el,\w{k})$ discussed in Sec.~\ref{s:normal_frame}.

The vectors $\el$ and $\w{k}$ are necessarily colinear to respectively
the vectors $\w{\tilde\ell}$ and $\w{\tilde k}$ associated with the 
dual-null foliation introduced above, i.e. there exists two positive scalar 
fields, $A$ and $B$, such that
\be \label{e:elk_tilde_elk}
    \el = A \w{\tilde\ell}
    \qquad\mbox{and}\qquad
    \w{k} = B \w{\tilde k} .
\ee 
Actually, we will use Eq.~(\ref{e:elk_tilde_elk}) to define $\el$
and $\w{k}$ away from $\Hor$, Eq.~(\ref{e:h_el_k}) defining them only 
on $\Hor$. However, all the results presented here are independent of the values
of $A$ and $B$ away from $\Hor$. The normalization $\el\cdot\w{k} = - 1$,
combined with Eq.~(\ref{e:tl_dot_tk}) relates the product $AB$ to $f$:
\be
    AB = e^{-f} \qquad\mbox{or}\qquad f = - \ln(AB) .
\ee
Then, from Eqs.~(\ref{e:elk_tilde_elk}) and (\ref{e:def_hat_lk}),
\be
    \el = B^{-1} \w{\hat\ell}
    \qquad\mbox{and}\qquad
    \w{k} = A^{-1} \w{\hat k} ,
\ee
which implies 
\be
    \w{h} \equalH B^{-1} \w{\hat\ell} - C  A^{-1} \w{\hat k} .
\ee
Consequently, taking into account Eqs.~(\ref{e:def_tilde_lk}),
(\ref{e:def_hat_lk}) and (\ref{e:lie_el_v}),
\be
    \Lie{\w{h}} u \equalH - CA^{-1}
    \qquad\mbox{and}\qquad
    \Lie{\w{h}} v \equalH B^{-1} . 
\ee
On the other side, since $u\equalH u_0(t)$ and
$v\equalH v_0(t)$, $\Lie{\w{h}} u \equalH {u'}_0(t)$ and
$\Lie{\w{h}} v \equalH {v'}_0(t)$, where we have used the fundamental
property $\Lie{\w{h}} t = 1$ defining $\w{h}$ [Eq.~(\ref{e:Lieh_t_1})].
We then conclude that 
\be
    B \equalH 1/{v'}_0(t)
    \qquad\mbox{and}\qquad
    C/A \equalH - {u'}_0(t) .
\ee
This implies that, on $\Hor$, the fields $B$ and $C/A$ are functions
of $t$ only; in particular, they are constant on each 2-surface $\Sp_t$:
\be \label{e:const_B_CsA}
    \DS B \equalH 0 
    \qquad\mbox{and}\qquad
    \DS (C/A) \equalH 0 .
\ee

Using Eqs.~(\ref{e:elk_tilde_elk}) and (\ref{e:dtl_dtk}), we get
\be \label{e:exterior_lk}
    \dd \underline{\el} = \dd \ln A \wedge 
     \underline{\w{\el}}
    \qquad \mbox{and} \qquad
    \dd \underline{\w{k}} = \dd \ln B \wedge 
     \underline{\w{k}} ,
\ee 
from which we obtain
\be \label{e:inaff_l}
    \w{\nabla}_{\el} \, \el = \nu_{(\el)} \, \el 
     \qquad \mbox{with} \qquad \nu_{(\el)} := \Lie{\el}\ln A   ,
\ee
\be \label{e:inaff_k}
    \w{\nabla}_{\w{k}} \, \w{k} = \nu_{(\w{k})} \, \w{k} 
     \qquad \mbox{with} \qquad \nu_{(\w{k})} := \Lie{\w{k}}\ln B  .
\ee
$\nu_{(\el)}$ and $\nu_{(\w{k})}$ are the \emph{inaffinity parameters}
of the null vector fields $\el$ and $\w{k}$.
Using the definitions (\ref{e:def_deform}), (\ref{e:def_Omega_l})
and (\ref{e:def_Omega_k}), we then get an expression for the
spacetime gradients of $\el$ and $\w{k}$:
\be  \w{\nabla} \underline{\el} = \w{\Theta}^{(\el)}
    + \underline{\el} \otimes \w{\Omega}^{(\el)}
    - \nu_{(\el)} \underline{\el} \otimes \underline{\w{k}}
    - \w{\nabla}_{\w{k}} \underline{\el} \otimes  \underline{\el}
            \label{e:grad_ul} 
\ee
\be
   \w{\nabla} \underline{\w{k}} = \w{\Theta}^{(\w{k})}
    - \underline{\w{k}} \otimes \w{\Omega}^{(\el)}
    - \nu_{(\w{k})} \underline{\w{k}} \otimes \underline{\el}
    - \w{\nabla}_{\el} \underline{\w{k}} \otimes  \underline{\w{k}} , 
            \label{e:grad_uk}
\ee
where we have used Eq.~(\ref{e:Omega_k_Omega_l}) with $\sigma=0$
to set $\w{\Omega}^{(\w{k})} = - \w{\Omega}^{(\el)}$. 
Besides, from Eqs.~(\ref{e:exterior_lk}) and (\ref{e:const_B_CsA}),
we get useful identities:
\bea
    \vqs \w{\nabla}_{\w{k}} \, \underline{\el} & = & -  \w{\Omega}^{(\el)}
        + \DS \ln A \label{e:grad_k_ul} \\
    \vqs \w{\nabla}_{\el} \, \underline{\w{k}} & = & \w{\Omega}^{(\el)} .
        \label{e:grad_l_uk}
\eea

\subsection{``Surface-gravity'' 1-forms}

Let us define the 1-form
\be \label{e:def_kappa_l}
\w{\kappa}^{(\el)} :=  \frac{1}{\w{k}\cdot\el} \, 
        \w{k} \cdot \w{\nabla}_{\w{\stackrel{\perp}{\vec{q}}}}\,  \el 
    \quad \left[\mbox{or}\ \kappa^{(\el)}_\alpha := \frac{1}{k_\rho \ell^\rho}
    k_\mu \nabla_\nu \ell^\mu \, 
        {\stackrel{\perp}{q}}^\nu_{\ \, \alpha} \right] ,
\ee
where $\w{\stackrel{\perp}{\vec{q}}}$ denotes the orthogonal projector
on the vector plane $\T_p(\Sp_t)^\perp$, i.e. the complementary of
$\w{\vec{q}}$: $\w{1} = \w{\vec{q}} \, + \w{\stackrel{\perp}{\vec{q}}}$.
The definition (\ref{e:def_kappa_l}) is similar to the 
definition (\ref{e:def_Omega_l}) of $\w{\Omega}^{(\el)}$, except for
$\w{\vec{q}}$ replaced by $\w{\stackrel{\perp}{\vec{q}}}$. Hence,
whereas $\w{\Omega}^{(\el)}$ was defined for a single 2-surface $\Sp_t$,
$\w{\kappa}^{(\el)}$ requires the knowledge of the null normal $\el$
in directions normal to $\Sp_t$.  From Eq.~(\ref{e:inaff_l})
and (\ref{e:inaff_k}), the inaffinity parameters $\nu_{(\el)}$
and $\nu_{(\w{k})}$ are recovered by applying the 1-form $\w{\kappa}^{(\el)}$
to respectively $\el$ and $-\w{k}$:
\be
    \langle \w{\kappa}^{(\el)}, \el \rangle = \nu_{(\el)}
    \qquad \mbox{and}\qquad
    \langle \w{\kappa}^{(\el)}, \w{k} \rangle = -\nu_{(\w{k})}  .  
\ee
A useful relation is then
\be \label{e:kappa_el_h}
    \langle \w{\kappa}^{(\el)}, \w{h} \rangle = \nu_{(\el)}
    + C \nu_{(\w{k})} .
\ee

\subsection{Trapping horizons and dynamical horizons} \label{s:def_mott}

Let us recall here the various definitions involved in the
local characterizations of black holes mentioned in the Introduction. 
The hypersurface $\Hor$ equipped with the spacelike foliation 
$(\Sp_t)_{t\in\mathbb{R}}$ is called a \emph{marginally outer trapped tube
(MOTT)} \cite{Booth05} if each leaf $\Sp_t$ is a marginal outer trapped surface
(cf. Sec.~\ref{s:normal_frame}), i.e. if $\theta^{\el}=0$ at any point 
in $\Hor$.  
Following Hayward \cite{Haywa94b} a \emph{trapping horizon}
is a MOTT on which $\theta^{(\w{k})} \not = 0$ and
$\Lie{k}\theta^{(\el)} \not = 0$, being qualified as \emph{future trapped
horizon} if $\theta^{(\w{k})}<0$ and  \emph{future outer trapped
horizon (FOTH)} if $\theta^{(\w{k})}<0$ and $\Lie{k}\theta^{(\el)} < 0$,
the latter subcase being the one relevant for black holes (see Ref.~\cite{Booth05}
for a discussion). The \emph{dynamical horizons} introduced by 
Ashtekar and Krishnan \cite{AshteK02,AshteK03,AshteK04} are MOTT
such that (i) $\Hor$ is spacelike and (ii) $\theta^{(\w{k})} < 0$. 
In particular, a spacelike future trapping horizon is a dynamical horizon. 
When it is null, a MOTT is called a \emph{non-expanding horizon}
\cite{Hajic73,Hajic75,AshteBL02}. It corresponds to a black hole in
equilibrium.

%%%%%%%%%%%%%%%%%%%%%%%%%%%%%%%%%%%%%%%%%%%%%%%%%%%%%%%%%%%%%%%%%%%%%%%%%%%%%%%

\section{The generalized Damour-Navier-Stokes equation} \label{s:DNS}

\subsection{Original Damour-Navier-Stokes equation}

In the case where the hypersurface $\Hor$ is null, and in particular
when $\Hor$ is the event horizon of a black hole, 
$\w{h}=\el$ and the Damour-Navier-Stokes 
equation \cite{Damou79,Damou82,ParikW98,GourgJ05} writes
\bea    
    \LieS{\el}  \w{\Omega}^{(\el)} + \theta^{(\el)} \w{\Omega}^{(\el)}
    & = & \DS \nu_{(\el)} 
    - \DS \cdot \vec{\w{\sigma}}^{(\el)}
    + \frac{1}{2} \DS \theta^{(\el)} \nonumber \\
   & &  + 8\pi \vqs \w{T}\cdot\el .   \label{e:DNS_ori}
\eea
This equation is derived from the Einstein equation, as the presence of
the stress-energy tensor $\w{T}$ testifies\footnote{We are using geometrized
units, in which both the speed of light $c$ and the gravitation constant
$G$ are set to $1$}. 
It has exactly the same structure as a 2-dimensional Navier-Stokes 
equation: 
dividing Eq.~(\ref{e:DNS_ori}) by $8\pi$, 
$-\w{\Omega}^{(\el)} /(8\pi)$ is interpreted by Damour \cite{Damou79,Damou82}
as a momentum surface density,  $\nu_{(\el)}/(8\pi)$ as a ``fluid''
pressure, $1/(16\pi)$ as a shear viscosity ($\w{\sigma}^{(\el)}$ being the
shear tensor), $-1/(16\pi)$ in front of $\DS \theta^{(\el)}$
as a bulk viscosity and
$\vqs \w{T}\cdot\el$ as a force surface density. 
The reader is referred to Chap.~VI of Ref.~\cite{ThornPM86} for an extended
discussion of this ``viscous fluid'' viewpoint. 

\subsection{Derivation of the generalized equation}

\begin{figure}
\includegraphics[width=0.3\textwidth]{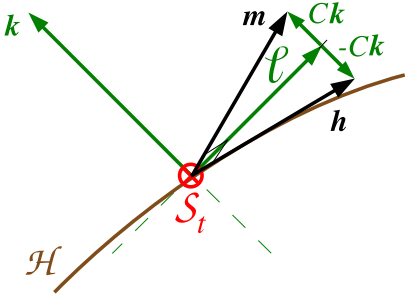} 
\caption{ \label{f:vector_m}
Same as Fig.~\ref{f:coupe_normal} but with in addition the 
vector $\w{m}$ normal to the hypersurface $\Hor$.}
\end{figure}

First of all, it must be noted that in the Damour-Navier-Stokes 
equation (\ref{e:DNS_ori}), 
the vector field $\el$ plays two different roles: 
it is both the evolution vector along
$\Hor$ (obviously in a term like $\LieS{\el}$) and 
the normal to $\Hor$ (in a term like $\vqs \w{T}\cdot\el$). 
When $\Hor$ is no longer null, these
two roles have to be taken by two different vectors. We have already
seen that the privileged evolution vector along $\Hor$ is the vector
$\w{h}$ associated with the foliation $(\Sp_t)_{t\in\mathbb{R}}$.
Regarding the normal vector, it is natural to consider
\be \label{e:def_m}
    \w{m} := \el + C \w{k} . 
\ee
Indeed this vector is normal to $\Hor$, since by construction 
$\w{m}\in\T_p(\Sp_t)^\perp$ and $\w{m}\cdot\w{h}=0$, and in the limit 
where $\Hor$ is null, it reduces to $\el$. It can be viewed as the unique
normal vector to $\Hor$ whose projection onto $\Hor$ along
the ingoing null direction $\w{k}$ is $\w{h}$
(see Fig.~\ref{f:vector_m}). Note that the scalar square of 
$\w{m}$ is the negative of that of $\w{h}$:
\be \label{e:norm_m}
    \w{m}\cdot\w{m} = - 2 C = - \w{h}\cdot\w{h} . 
\ee

A generalization of the Damour-Navier-Stokes equation to the non-null case
should contain the term $\vqs \w{T}\cdot\w{m}$, instead of
$\vqs\w{T}\cdot\el$, in the right-hand side. By virtue of Einstein
equation and the fact that $\vec{\w{q}}(\w{m})=0$, 
$8\pi \vqs\w{T}\cdot\w{m} = \vqs \w{R}\cdot\w{m}$, where $\w{R}$ is
the Ricci tensor associated with the spacetime metric $\w{g}$. The starting
point for getting the generalized Damour-Navier-Stokes equation will
be then the contracted Ricci identity applied to the vector $\w{m}$
and projected onto $\Sp_t$:
\be \label{e:Ricci_m}
    \left( \nabla_\mu \nabla_\nu m^\mu 
    - \nabla_\nu \nabla_\mu m^\mu \right) q^\nu_{\ \, \alpha}  = 
    R_{\mu\nu} m^\mu q^\nu_{\ \, \alpha} .
\ee 
Now, by combining definition (\ref{e:def_m}) with expressions
(\ref{e:grad_ul}) and (\ref{e:grad_uk}),
\bea
    \nabla_\alpha m^\beta & = & \Theta_\alpha^{(\w{m})\beta}
    + \Omega_\alpha^{(\el)} h^\beta - \nu_{(\el)} k_\alpha \ell^\beta
    - C \nu_{(\w{k})} \ell_\alpha k^\beta \nonumber \\
    & & - \ell_\alpha k^\sigma \nabla_\sigma \ell^\beta 
    - C k_\alpha \ell^\sigma \nabla_\sigma k^\beta 
    + \nabla_\alpha C k^\beta . \label{e:grad_m}
\eea
Substituting Eq.~(\ref{e:grad_m}) for $\nabla_\nu m^\mu$ and
$\nabla_\mu m^\mu$ in Eq.~(\ref{e:Ricci_m}), expanding and making
use of identities (\ref{e:grad_k_ul}), (\ref{e:grad_l_uk})
and (\ref{e:kappa_el_h})
yields, after some rearrangements,
\begin{widetext}
\bea
    R_{\mu\nu} m^\mu q^\nu_{\ \, \alpha} & = & 
      q^\nu_{\ \, \alpha}   \nabla_\mu  \Theta_\nu^{(\w{m})\mu} + 
      q^\nu_{\ \, \alpha}  h^\mu \nabla_\mu  \Omega_\nu^{(\el)} 
      + \Theta_\alpha^{(\w{h})\mu} \, \Omega_\mu^{(\el)}
      + \theta^{(\w{h})} \, \Omega_\alpha^{(\el)} 
      - \DSc_\alpha \left( \theta^{(\w{m})} 
      + \langle \w{\kappa}^{(\el)}, \w{h} \rangle  \right)
      - \Theta_\alpha^{(\el)\mu} \, \DSc_\mu \ln A     
       \nonumber \\
       & & -  \Theta_\alpha^{(\w{k})\mu} \, \DSc_\mu C
       + \theta^{(\w{k})} \DSc_\alpha C 
        +  \nu_{(\w{k})} 
  \underbrace{\left( \DSc_\alpha C - C \DSc_\alpha \ln A \right)}_{=0} ,
    \label{e:exDNS_prov}
\eea 
where the ``$=0$'' results from Eq.~(\ref{e:const_B_CsA}). 
Now, from the relation (\ref{e:DS_nabla}) between the derivatives
$\w{\DS}$ and $\w{\nabla}$, one has [making use of identities 
(\ref{e:grad_k_ul}) and (\ref{e:grad_l_uk})]
\be \label{e:q_nab_Theta_m}
    q^\nu_{\ \, \alpha} \nabla_\mu  \Theta_\nu^{(\w{m})\mu}
    = \DSc_\mu  \Theta_\alpha^{(\w{m})\mu} + 
    \Theta_\alpha^{(\w{m})\mu} \, \DSc_\mu \ln A .  
\ee
Besides, expressing the Lie derivative in terms of $\w{\nabla}$ gives
\be \label{e:q_Lie_Omega_l}
  q^\nu_{\ \, \alpha} \Liec{\w{h}} \Omega_\nu^{(\el)} =
    q^\nu_{\ \, \alpha} h^\mu \nabla_\mu  \Omega_\nu^{(\el)} 
    +  \Theta_\alpha^{(\w{h})\mu} \, \Omega_\mu^{(\el)} . 
\ee
Thanks to Eqs.~(\ref{e:q_nab_Theta_m}) and (\ref{e:q_Lie_Omega_l}),
Eq.~(\ref{e:exDNS_prov}) reduces to 
\be
   R_{\mu\nu} m^\mu q^\nu_{\ \, \alpha} = 
     \DSc_\mu  \Theta_\alpha^{(\w{m})\mu}
     + q^\nu_{\ \, \alpha} \Liec{\w{h}} \Omega_\nu^{(\el)} 
     + \theta^{(\w{h})}  \, \Omega_\alpha^{(\el)} 
      - \DSc_\alpha \left( \theta^{(\w{m})} + 
      \langle \w{\kappa}^{(\el)}, \w{h} \rangle
      \right)  + \theta^{(\w{k})} \DSc_\alpha C ,
\ee
where Eq.~(\ref{e:const_B_CsA}) has been used to set to zero
the term $\Theta_\alpha^{(\w{k})\mu} \left(C \DSc_\mu \ln A  - \DSc_\mu C
\right)$ which had appeared. 
Now, from Eq.~(\ref{e:LieS_Lie}) and the property
$\vqs \w{\Omega}^{(\el)}= \w{\Omega}^{(\el)}$, the term 
$\vqs \Lie{\w{h}} \w{\Omega}^{(\el)}$ which appears in the above equation is
nothing but $\LieS{\w{h}} \w{\Omega}^{(\el)}$. Re-expressing 
$\w{\Theta}^{(\w{m})}$ in terms
of the shear tensor $\w{\sigma}^{(\w{m})}$ and the expansion scalar
$\theta^{(\w{m})}$ via Eq.~(\ref{e:def_shear}) and taking account the 
Einstein equation then leads to 
\be
    \LieS{\w{h}} \w{\Omega}^{(\el)} + \theta^{(\w{h})} \, \w{\Omega}^{(\el)}
    = \DS \langle \w{\kappa}^{(\el)}, \w{h} \rangle
        - \DS \cdot \vec{\w{\sigma}}^{(\w{m})}
             + \frac{1}{2} \DS \theta^{(\w{m})}
      - \theta^{(\w{k})} \DS C 
    + 8\pi \vqs \w{T}\cdot \w{m} .   \label{e:genDNS}
\ee
\end{widetext}
This is the generalization of Damour-Navier-Stokes equation to the case
where the foliated hypersurface $\Hor$ is not necessarily null. In the
null limit, $C=0$, $\w{h}=\w{m}=\el$ and we recover Damour's original
version, i.e. Eq.~(\ref{e:DNS_ori}). In the non-null case, it is 
worth to notice that the obtained equation is not much more complicated
than Eq.~(\ref{e:DNS_ori}): apart from substitutions of $\el$ by
either $\w{h}$ or $\w{m}$, as discussed above, it contains only one 
extra term: $\theta^{(\w{k})} \DS C$.

\subsection{Change of normal fundamental form}

Let us rewrite the generalized Damour-Navier-Stokes equation in 
terms of the normal fundamental form $\w{\Omega}^{(\el')}$ associated with
a generic null vector $\el'$, instead of $\el$. 
Setting $\w{\el}' = \lambda\el$, with $\lambda>0$,
$\w{\Omega}^{(\el')}$ is related to 
$\w{\Omega}^{(\el)}$ by Eq.~(\ref{e:rescale_Omega_l}), from which 
we deduce
\bea
    \LieS{\w{h}} \w{\Omega}^{(\el')} & + & 
    \theta^{(\w{h})} \, \w{\Omega}^{(\el')}
     = \LieS{\w{h}} \w{\Omega}^{(\el)} + 
     \theta^{(\w{h})} \, \w{\Omega}^{(\el)} \nonumber \\
      & &+ \DS \left( \Lie{\w{h}} \ln\lambda \right)
     + \theta^{(\w{h})}  \DS \ln\lambda ,   \label{e:Lie_Omega_l_prim}
\eea
where we have used 
$\LieS{\w{h}} \DS \ln\lambda = \DS (\Lie{\w{h}} \ln\lambda )$.
Besides, we note that the 1-form $\w{\kappa}^{(\el)}$ transforms
as follows
\be
   \w{\kappa}^{(\el')} = \w{\kappa}^{(\el)} + 
    \w{\nabla}_{\w{\stackrel{\perp}{\vec{q}}}} \ln\lambda, 
\ee
from which
\be
    \DS \langle \w{\kappa}^{(\el')}, \w{h} \rangle = 
    \DS \langle \w{\kappa}^{(\el)}, \w{h} \rangle 
    + \DS\left( \Lie{\w{h}} \ln\lambda \right) . \label{e:dkappa_l_prim}
\ee
Combining Eqs.~(\ref{e:genDNS}), (\ref{e:Lie_Omega_l_prim})
and (\ref{e:dkappa_l_prim}), and using
$\theta^{(\w{h})} = \theta^{(\el)} - C \theta^{(\w{k})}$
yields
\begin{widetext}
\be
    \LieS{\w{h}} \w{\Omega}^{(\el')} + \theta^{(\w{h})} \, \w{\Omega}^{(\el')}
    = \DS \langle \w{\kappa}^{(\el')}, \w{h} \rangle
        - \DS \cdot \vec{\w{\sigma}}^{(\w{m})}
             + \frac{1}{2} \DS \theta^{(\w{m})}
      + \theta^{(\el)}  \DS \ln\lambda
      -  \theta^{(\w{k})} \left( \DS C +  C \DS \ln\lambda \right)  
    + 8\pi \vqs \w{T}\cdot \w{m} .    \label{e:genDNS_el_prim}
\ee
If we chose $\lambda=A^{-1}$, then by virtue of Eq.~(\ref{e:const_B_CsA}),
the term $\DS C +  C \DS \ln\lambda$ vanishes identically. Since 
$\lambda=A^{-1}$ corresponds to $\el'=\w{\tilde\ell}$ 
[cf. Eq.~(\ref{e:elk_tilde_elk})], Eq.~(\ref{e:genDNS_el_prim}) reduces
then to 
\be
    \LieS{\w{h}} \w{\Omega}^{(\w{\tilde\ell})} 
    + \theta^{(\w{h})} \, \w{\Omega}^{(\w{\tilde\ell})}
    = \DS \langle \w{\kappa}^{(\w{\tilde\ell})}, \w{h} \rangle
        - \DS \cdot \vec{\w{\sigma}}^{(\w{m})}
             + \frac{1}{2} \DS \theta^{(\w{m})}
      - \theta^{(\el)}  \DS \ln A
    + 8\pi \vqs \w{T}\cdot \w{m} .    \label{e:genDNS_tel}
\ee
\end{widetext}
In the case where $\Hor$ is not null ($C\not=0$), 
another way to set to zero the term $\DS C +  C \DS \ln\lambda$ in
Eq.~(\ref{e:genDNS_el_prim}) is to choose
\be
    \lambda = F(t) / |C| , 
\ee
where $F(t)$ is an arbitrary function of $t$,
since then $\DS C +  C \DS \ln\lambda = C \DS\ln(\lambda |C|)=C\DS\ln F(t)=0$. 
The simplest choice $F(t)=1$ corresponds to the following 
decomposition of the evolution vector:
$\w{h} = \pm( C \el' - \w{k}')$ (with $\pm$ corresponding 
to $C>0$ and $C<0$ respectively), to be contrasted with the decomposition
(\ref{e:h_el_k}). Actually this amounts simply to swapping the
vectors $\el$ and $\w{k}$.

\subsection{Application to trapping horizons}

If $\Hor$ is a trapping horizon (or more generally a MOTT, 
cf. Sec.~\ref{s:def_mott}), then $\theta^{(\el)}=0$ and
Eq.~(\ref{e:genDNS_tel}) becomes
\bea
     & &   \LieS{\w{h}} \w{\Omega}^{(\w{\tilde\ell})} 
    + \theta^{(\w{h})} \, \w{\Omega}^{(\w{\tilde\ell})}
    = \DS \langle \w{\kappa}^{(\w{\tilde\ell})}, \w{h} \rangle
        - \DS \cdot \vec{\w{\sigma}}^{(\w{m})} \nonumber \\
      & & \ \quad + \frac{1}{2} \DS \theta^{(\w{m})}       
    + 8\pi \vqs \w{T}\cdot \w{m} .    \label{e:genDNS_fth}
\eea
This equation is structurally identical to the original Damour-Navier-Stokes
equation [Eq.~(\ref{e:DNS_ori})]: apart from substitutions of $\el$ by
either $\w{h}$ or $\w{m}$, it does not contain any extra term.
The differences are that the original Damour-Navier-Stokes applies
to a null $\Hor$ but with $\theta^{(\el)}$ not necessarily zero,
whereas Eq.~(\ref{e:genDNS_fth}) is valid for both $\Hor$ null or 
spacelike, but assumes $\theta^{(\el)}=0$. 

%%%%%%%%%%%%%%%%%%%%%%%%%%%%%%%%%%%%%%%%%%%%%%%%%%%%%%%%%%%%%%%%%%%%%%%%%%%%%%%

\section{Angular momentum}
\label{s:angu_mom}

Traditionally the concept of \emph{angular momentum} is a global one
and requires the evaluation of a Komar integral at spatial infinity,
assuming $\M$ to be asymptotically flat and endowed with 
an axisymmetric Killing vector (cf. e.g. Ref.~\cite{Poiss04}).
However, by means of some Hamiltonian analysis, the concept
of angular momentum can be made quasilocal, as a quantity associated
with the interior of the spacetime region delimited by the hypersurface $\Hor$. 
The prototype of such quasilocal formulation is Brown-York analysis
\cite{BrownY93} which will be taken as the starting point for
our discussion. 

\subsection{Brown-York angular momentum}

Let us assume that the hypersurface $\Hor$ is timelike and is axisymmetric,
with the associated Killing vector $\w{\varphi}$ lying in the 2-surfaces
$\Sp_t$. The definition of angular momentum by Brown and York 
\cite{BrownY93} is then  
\be \label{e:def_J_BY}
    J := \oint_{\Sp_t} 
    \langle \w{j}, \w{\varphi} \rangle \, \volS ,
\ee
where $\volS$ is the surface element of $\Sp_t$ associated with the
induced metric $\w{q}$ ($\volS = \sqrt{q}\,  \dd x^2 \wedge \dd x^3$
for any coordinate system $x^a=(x^2,x^3)$ on $\Sp_t$,
with $q:=\det q_{ab}$)
and the momentum surface density 1-form $\w{j}$ 
is defined as follows. 
Adopting a 3+1 perspective (cf. Sec.~\ref{s:perspective}), let $\Sigma_t$
be a spacelike hypersurface intersecting $\Hor$ in $\Sp_t$. Denoting by
$\w{\gamma}$ and $\w{K}$ the induced metric and extrinsic curvature
tensor of $\Sigma_t$, $\w{j}$ is expressible as 
\be \label{e:def_jBY}
    j_\alpha = - \frac{2}{\sqrt{\gamma}} P^{\mu\nu} s_\mu q_{\nu\alpha} , 
\ee
where $\w{s}$ is the unit spacelike normal to $\Sp_t$ which lies in 
$\Sigma_t$ and $\w{P}$ is the momentum canonically conjugate to $\w{\gamma}$:
\be \label{e:def_mom_conjugate}
    P^{\alpha\beta} = \frac{1}{16\pi} \sqrt{\gamma}
    \left( K \gamma^{\alpha\beta}
    - K^{\alpha\beta} \right) . 
\ee
Since $\w{K}$ is related to the gradient of the timelike unit normal
to $\Sigma_t$, $\w{n}$, by $K_{\alpha\beta} = - \nabla_\mu n_\nu
\, \gamma^\mu_{\ \, \alpha} \gamma^\nu_{\ \, \beta}$, we get, by inserting
Eq.~(\ref{e:def_mom_conjugate}) in Eq.~(\ref{e:def_jBY}) and
comparing with Eq.~(\ref{e:def_Omega_n}),
\be
    \w{j} = - \frac{1}{8\pi} \w{\Omega}^{(\w{n})} . 
\ee  
By considering the null vector $\el':=\w{n}+\w{s}$ and combining the
transformation laws (\ref{e:Omega_el_n}) and (\ref{e:rescale_Omega_l}),
one gets
\be \label{e:jBY_Omega_el}
    \w{j} = - \frac{1}{8\pi} \left( \w{\Omega}^{(\el)}
        + \DS \ln \lambda \right) , 
\ee
where $\lambda$ is the scale factor relating $\el$ to $\w{n}+\w{s}$:
$\w{n}+\w{s} = \lambda \el$. Now, the Killing equation for the
vector $\w{\varphi}$
and the fact that $\w{\varphi}\in\T(\Sp_t)$ imply 
$\DS\cdot \w{\varphi}=0$. Consequently 
$\w{\varphi}\cdot\DS \ln \lambda = \DS\cdot(\ln\lambda\, \w{\varphi})$
is a perfect divergence, the integral of which over the closed surface
$\Sp_t$ vanishes. Therefore substituting Eq.~(\ref{e:jBY_Omega_el}) 
for $\w{j}$ into
Eq.~(\ref{e:def_J_BY}) yields
\be \label{e:J_BY_Omega_l}
    J = - \frac{1}{8\pi} \oint_{\Sp_t} 
    \langle \w{\Omega}^{(\el)}, \w{\varphi} \rangle \, \volS . 
\ee
This expression is in perfect agreement with the interpretation of
$-\w{\Omega}^{(\el)}/(8\pi)$ as a momentum surface density performed
in Sec.~\ref{s:DNS}.

\subsection{Generalized angular momentum}

It may be noticed that the timelike character of the hypersurface
$\Hor$, assumed in 
Brown-York Hamiltonian analysis \cite{BrownY93}, does not play 
any role in the expression (\ref{e:J_BY_Omega_l}) of the angular momentum. 
Actually the definition of angular momentum, based on 
Eq.~(\ref{e:J_BY_Omega_l}), has been extended to null hypersurfaces
by Booth \cite{Booth01} (in a generalization of Brown-York analysis)
and Ashtekar et al. \cite{AshteBL01} (in the framework of isolated
horizons). 

It is also worth to notice that the independence of the integral defining
$J$ with respect to the normal fundamental form (i.e. $\w{\Omega}^{(\w{n})}$
or $\w{\Omega}^{(\el)}$) stems only from the divergence-free property
of the vector $\w{\varphi}$, which is a condition weaker than that of
being a Killing vector. Therefore, one may relax the later and 
follow Booth and Fairhurst's recent analysis \cite{BoothF05}
to introduce a generalized angular momentum as follows. Let us  
assume that the 2-surfaces $\Sp_t$ have the topology
of $\mathbb{S}^2$. 
Let $\w{\varphi}$ be a vector field in $\T(\Sp_t)$ which (i)
has closed orbits and (ii) has vanishing divergence with respect to 
the induced connection $\DS$: 
\be \label{e:div_varphi_zero}
    \DS\cdot\w{\varphi} = 0 .  
\ee
The \emph{angular momentum associated with $\w{\varphi}$} is then
defined by \cite{BoothF05} 
\be \label{e:def_genJ}
    J(\w{\varphi}) := - \frac{1}{8\pi} \oint_{\Sp_t} 
    \langle \w{\Omega}^{(\el)}, \w{\varphi} \rangle \, \volS , 
\ee
which is a formula structurally identical to formula (\ref{e:J_BY_Omega_l}).
The main difference is that $\w{\varphi}$ is no longer the Killing vector
reflecting the axisymmetry of $\Sp_t$ and uniquely defined by the normalization
of the orbit lengths to $2\pi$, but merely a divergence-free
vector field. Consequently, $J$ depends on the choice of $\w{\varphi}$.
However formula (\ref{e:def_genJ}) shares with formula 
(\ref{e:J_BY_Omega_l}) the independence with respect to 
the choice of the normal fundamental form $\w{\Omega}^{(\el)}$,
thanks to the divergence-free character of $\w{\varphi}$. 
Indeed, under a change of null 
normal $\el'=\lambda\el$, $\w{\Omega}^{(\el)}$ is changed
to $\w{\Omega}^{(\el')} = \w{\Omega}^{(\el)} + \DS \ln \lambda$
[Eq.~(\ref{e:rescale_Omega_l})] and since $\DS\cdot\w{\varphi}=0$,
$\w{\varphi}\cdot\DS \ln \lambda = \DS\cdot(\ln\lambda\, \w{\varphi})$
is a perfect divergence, the integral of which on $\Sp_t$ vanishes. 

As a further justification of definition (\ref{e:def_genJ}), it
is shown in Appendix~\ref{s:AK} that, when $\Hor$ is a dynamical horizon, 
$J(\w{\varphi})$ agrees with the generalized angular momentum 
defined  by Ashtekar and Krishnan \cite{AshteK03,AshteK04}.

\subsection{Angular momentum flux law}

For any 1-form $\w{\omega}$ and vector field $\w{\varphi}$
defined on $\Hor$ and both tangent to $\Sp_t$ for
all $t\in\mathbb{R}$, the following identity holds:
\bea
\frac{d}{dt} \oint_{\Sp_t} 
    \langle \w{\omega}, \w{\varphi} \rangle \, \volS &=& 
    \oint_{\Sp_t} \LieS{h} 
        \! \left[ \langle \w{\omega}, \w{\varphi} \rangle \, \volS \right]
        \nonumber \\
    &=& \oint_{\Sp_t} \langle \LieS{h} \w{\omega}
    + \theta^{(\w{h})} \w{\omega}, \w{\varphi} \rangle \volS \nonumber \\
   & & + \oint_{\Sp_t} \langle \w{\omega}, \Lie{h} \w{\varphi} \rangle 
    \volS , 
\eea
where the first equality results from the Lie transport of the 2-surfaces
$\Sp_t$ by the vector field $\w{h}$ (cf. Sec.~\ref{s:evol_vect})
and in the second equality the relation
$\Lie{h} \volS = \theta^{(\w{h})}\volS$ has been used 
[cf. Eq.~(\ref{e:Liev_volS})].

Applying the above identity to the 1-form $\w{\omega} = \w{\Omega}^{(\el)}$
and employing the generalized Damour-Navier-Stokes equation 
(\ref{e:genDNS}) leads to an evolution equation for the 
generalized angular momentum
$J(\w{\varphi})$ defined by Eq.~(\ref{e:def_genJ}): 
\begin{widetext}
\be \label{e:Jflux_prov}
    \frac{d}{dt} J(\w{\varphi}) = - \oint_{\Sp_t} \w{T}(\w{m},\w{\varphi})
    \, \volS
    - \frac{1}{16\pi} \oint_{\Sp_t} \left[ 
  \stackrel{\twoheadrightarrow}{\w{\sigma}}\!\!{}^{(\w{m})}\!:\Lie{\varphi}\w{q}
       - 2 \theta^{(\w{k})} \w{\varphi}\cdot \DS C \right]
        \volS 
    - \frac{1}{8\pi}
     \oint_{\Sp_t} \langle \w{\Omega}^{(\el)}, \Lie{h} \w{\varphi} \rangle 
    \volS .
\ee
The notation `$:$' stands for a complete contraction, whereas the double arrow
means that the two indices of $\w{\sigma}^{(\w{m})}$ have been 
raised with the metric $\w{g}$: 
$\stackrel{\twoheadrightarrow}{\w{\sigma}}\!\!{}^{(\w{m})}\!:\Lie{\varphi}\w{q}
 = \sigma^{(\w{m})ab} \Liec{\varphi} q_{ab}$.
The integrals involving the pure gradients 
$\DS \langle \kappa^{(\el)}, \w{h} \rangle$ and 
$\DS \theta^{(\w{m})}$ have been set to zero thanks to the property
$\DS\cdot\w{\varphi}=0$. Besides, we have written
$\w{\varphi} \cdot \left( \DS \cdot \vec{\w{\sigma}}^{(\w{m})} \right)
= \DS \cdot \left( \vec{\w{\sigma}}^{(\w{m})}  \cdot \w{\varphi} \right) 
-  \vec{\w{\sigma}}^{(\w{m})} : \DS\w{\varphi}$, with the integral
of the divergence
$\DS \cdot \left( \vec{\w{\sigma}}^{(\w{m})}  \cdot \w{\varphi} \right)$
being zero since $\Sp_t$ is a closed manifold
and, thanks to the symmetry 
of the shear tensor $\w{\sigma}^{(\w{m})}$,
$2\vec{\w{\sigma}}^{(\w{m})}: \DS\w{\varphi}
= \sigma^{(\w{m})ab} (\DSc_a\varphi_b + \DSc_b \varphi_a )
= \stackrel{\twoheadrightarrow}{\w{\sigma}}\!\!{}^{(\w{m})}\!:\Lie{\varphi}\w{q}$.

The last integral in Eq.~(\ref{e:Jflux_prov}) occurs to take into account
a possible variation of $\w{\varphi}$ along the evolution vector $\w{h}$.
To make the variation of $J$ more meaningful, it is natural to 
demand that the vector field $\w{\varphi}$ is transported by 
$\w{h}$:
\be \label{e:varphi_htransp}
    \Lie{h}{\w{\varphi}} = 0 . 
\ee
From now on, we assume that $\w{\varphi}$ obeys to both conditions
(\ref{e:div_varphi_zero}) and (\ref{e:varphi_htransp}). Note that if
$\w{\varphi}$ is a symmetry generator of $\Hor$ which is tangent to
$\Sp_t$, these two conditions are satisfied (in particular
$\Lie{h} \w{\varphi} = - \Lie{\varphi} \w{h} = 0$).
Then Eq.~(\ref{e:Jflux_prov}) simplifies to
\be \label{e:Jflux}
    \frac{d}{dt} J(\w{\varphi}) = - \oint_{\Sp_t} \w{T}(\w{m},\w{\varphi})
    \, \volS
    - \frac{1}{16\pi} \oint_{\Sp_t}
        \left[ 
    \stackrel{\twoheadrightarrow}{\w{\sigma}}\!\!{}^{(\w{m})}\!:\Lie{\varphi}\w{q}
         - 2 \theta^{(\w{k})} \w{\varphi}\cdot \DS C \right]
        \volS  .
\ee
\end{widetext}

In the case where $\Hor$ is a null hypersurface, then $C=0$ and
Eq.~(\ref{e:Jflux}) reduces to 
\be \label{e:Jflux_null}
    \frac{d}{dt} J(\w{\varphi}) =
    - \oint_{\Sp_t} \w{T}(\el,\w{\varphi}) \, \volS 
     - \frac{1}{16\pi} \oint_{\Sp_t}
   \stackrel{\twoheadrightarrow}{\w{\sigma}}\!\!{}^{(\el)}\!:\Lie{\varphi}\w{q} 
        \; \volS  .
\ee
We recover here Eq.~(6.134) of the \emph{Membrane Paradigm} book 
\cite{ThornPM86}, where the first term is interpreted as the 
variation of $J$ due to the flux of angular momentum carried by matter
and electromagnetic field at $\Hor$ and the second term accounts for the
shear viscosity of $\Hor$.  

Let us consider now the case where $\Hor$ is either timelike or
spacelike: $C\not =0$ and we may express $\theta^{(\w{k})}$ in 
Eq.~(\ref{e:Jflux}) as 
$\theta^{(\w{k})} = \left( \theta^{(\el)} - \theta^{(\w{h})}
        \right) / C$ [cf. Eq.~(\ref{e:h_el_k})], so that  
\be \label{e:theta_k_DC}
    -\theta^{(\w{k})} \w{\varphi}\cdot \DS C = 
        - \theta^{(\el)} \w{\varphi}\cdot \DS \ln |C|
        +  \theta^{(\w{h})} 
        \w{\varphi}\cdot \DS \ln |C|.
\ee
Now, the properties (\ref{e:div_varphi_zero})
and (\ref{e:varphi_htransp}) fulfilled by $\w{\varphi}$ imply that
the vector field $\theta^{(\w{h})}  \w{\varphi}$ is divergence-free on
$(\Sp_t,\w{q})$: 
\be \label{e:th_vp_divfree}
    \DS\cdot \left( \theta^{(\w{h})}  \w{\varphi} \right) = 0 . 
\ee
This follows from the identity
\be
    \Lie{h} \DS\cdot\w{\varphi}
    + \theta^{(\w{h})} \DS\cdot\w{\varphi}
    = \DS\cdot \left( \Lie{h}\w{\varphi}
    + \theta^{(\w{h})}  \w{\varphi} \right) , 
\ee
which can be easily established, for instance by considering
a coordinate system $(t,x^2,x^3)$ on $\Hor$ such that 
$\w{h}=\partial/\partial t$. 
As a consequence of Eq.~(\ref{e:th_vp_divfree}), 
the integral over $\Sp_t$ of 
$\theta^{(\w{h})} \w{\varphi}\cdot \DS \ln |C|$ vanishes. 
Taking into account Eq.~(\ref{e:theta_k_DC}), we deduce then that 
Eq.~(\ref{e:Jflux}) can be written  
\begin{widetext}
\be
    \frac{d}{dt} J(\w{\varphi})  =
    - \oint_{\Sp_t} \w{T}(\w{m},\w{\varphi}) \volS 
     - \frac{1}{16\pi} \oint_{\Sp_t} \left[
     \stackrel{\twoheadrightarrow}{\w{\sigma}}\!\!{}^{(\w{m})}\!:\Lie{\varphi}\w{q} 
       - 2 \theta^{(\el)} \w{\varphi}\cdot \DS \ln |C| \right]
        \; \volS  . \label{e:Jflux_non_null}
\ee
\end{widetext}
We deduce immediately from this expression that if $\Hor$ is a 
timelike or spacelike MOTT, i.e. if $\theta^{(\el)}=0$, the angular momentum
variation law takes a very simple form:
\be \label{e:Jflux_trap}
    \frac{d}{dt} J(\w{\varphi}) =
    - \oint_{\Sp_t} \w{T}(\w{m},\w{\varphi}) \volS 
     - \frac{1}{16\pi} \oint_{\Sp_t}
     \stackrel{\twoheadrightarrow}{\w{\sigma}}\!\!{}^{(\w{m})}\!:\Lie{\varphi}\w{q} 
        \; \volS  .
\ee
In particular, the above formula holds for a spacelike future outer trapping
horizon\footnote{If the null energy condition holds, a non-null 
future outer trapping horizon is necessarily spacelike \cite{Haywa94b}.} 
and for a dynamical horizon. 
We have established this relation by assuming $C \not =0$.
Now the relation obtained in the null case, Eq.~(\ref{e:Jflux_null}), is
identical to Eq.~(\ref{e:Jflux_trap}) since $\w{m}=\el$ when $C=0$. 
Therefore, collecting the two results, we
conclude that Eq.~(\ref{e:Jflux_trap}) holds for a future outer trapping 
horizon of any kind: dynamical horizon ($C>0$) or non-expanding horizon
($C=0$).  

It is worth to note that Eq.~(\ref{e:Jflux_trap}) has
the same form as Eq.~(\ref{e:Jflux_null}), which has been established
for generic null hypersurfaces, not necessarily non-expanding horizons.  
In the case where $\Hor$ is a dynamical horizon, 
$\w{m}$ is timelike and since 
$\w{\varphi}$ is orthogonal to $\w{m}$, 
$-\w{T}(\w{m},\w{\varphi})$ represents a momentum density along
the spatial direction $\w{\varphi}$. Therefore we can attribute the
first term in right-hand side of Eq.~(\ref{e:Jflux_trap}) to the
flux of matter and electromagnetic angular momentum across $\Sp_t$. 
Regarding the second term, it clearly vanishes if $\Sp_t$
is axisymmetric, with $\w{\varphi}$ as a  
symmetry generator ($\Lie{\varphi}\w{q}=0$).  

\subsection{Relation to previous angular momentum laws}

Booth and Fairhurst \cite{BoothF04} have obtained the angular momentum
flux law (\ref{e:Jflux_trap}) is the case where $\Hor$ is a \emph{slowly
evolving horizon}.
A slowly evolving horizon is a future outer trapping
horizon which is close (in a sense made precise in Ref.~\cite{BoothF04})
to an isolated horizon. Eq.~(\ref{e:Jflux_trap}) is then obtained 
by an expansion to second order in $\epsilon$, where 
$\epsilon\simeq\sqrt{2C}$ is the small parameter which measures the deviation 
from an isolated horizon\footnote{Note that the
definition of angular momentum in Ref.~\cite{BoothF04}
has the opposite sign than ours.}. 
At this order, note that 
$\stackrel{\twoheadrightarrow}{\w{\sigma}}\!\!{}^{(\w{m})}$ is replaced
by $\stackrel{\twoheadrightarrow}{\w{\sigma}}\!\!{}^{(\el)}$ in 
Booth and Fairhurst's version [their Eq.~(14)]. 
Another difference with these Authors is that they did not assume
that $\w{\varphi}$ is divergence-free, but that it is close to a Killing
vector of $(\Sp_t,\w{q})$. 

In the case where $\Hor$ is a dynamical horizon, 
Ashtekar and Krishnan \cite{AshteK03} have 
also derived an angular momentum balance law, but in a time-integrated
form, so that it involves 3-dimensional integrals. 
Actually the relation derived in Ref.~\cite{AshteK03} does not assume 
that $\Hor$ is a MOTT and is valid for any spacelike hypersurface.  
It turns out that if we integrate with respect to $t$ the angular
momentum law (\ref{e:Jflux_non_null}), which is also valid for any spacelike
$\Hor$, we recover exactly Ashtekar and Krishnan's version. 
This is shown in Appendix~\ref{s:AK}. 

%%%%%%%%%%%%%%%%%%%%%%%%%%%%%%%%%%%%%%%%%%%%%%%%%%%%%%%%%%%%%%%%%%%%%%%%%%%%%%%

\section{Conclusion}
\label{s:concl}

We have established, by means of Einstein equation, an identity 
valid for any hypersurface $\Hor$ foliated by spacelike 2-surfaces
$(\Sp_t)_{t\in\mathbb{R}}$. 
This equation has the same form, up to some additional term, 
as the 2-dimensional Navier-Stokes
equation obtained by Damour \cite{Damou79,Damou82}
for describing the dynamics of event horizons. 
The evolution vector, giving the time derivative of the effective
momentum surface density is the vector $\w{h}$ tangent to $\Hor$, 
orthogonal to  the leaves $\Sp_t$ and which transports them into each other 
(Lie-dragging).
The role of the momentum surface density is played by the normal
fundamental form $\w{\Omega}^{(\el)}$ of $\Sp_t$
associated with the outgoing null normal $\el$ whose 
projection along the ingoing null direction is $\w{h}$.
The pressure term involves both vectors $\el$ and $\w{h}$, as it
is the ``surface gravity'' 1-form of $\el$ acting on $\w{h}$.  
The vector defining the shear and the expansion
involved in the viscous terms is the vector $\w{m}$ normal to $\Hor$
and whose projection onto $\Hor$
along the ingoing normal null direction is $\w{h}$. 
The vector $\w{m}$ gives also the external force exercised by matter
and electromagnetic fields, if any. 

It must be noted that another outgoing null vector can be selected
instead of $\el$, such as a tangent
$\w{\tilde\ell}$ to the hypersurfaces of outgoing light rays 
emanating orthogonally from $\Sp_t$, the dual 1-form of which is closed:
$\dd \underline{\w{\tilde\ell}}=0$. The key point is that all
normal fundamental forms and ``surface gravity'' 1-forms 
differ only by a gradient and their
interchange alters only slightly the generalized Damour-Navier-Stokes
equation.   

If the hypersurface $\Hor$ is null, the three vectors
$\el$, $\w{h}$ and $\w{m}$ coincide and the equation obtained here reduces
to the original Damour-Navier-Stokes equation \cite{Damou79,Damou82}. 
If $\Hor$ is a marginally outer trapped tube,
and in particular if it is a dynamical horizon or a future outer trapping
horizon, the obtained equation, written in terms of $\w{\tilde\ell}$, 
is as simple as the original Damour-Navier-Stokes
equation, the generic additional term vanishing in this case. 
 
The generalized Damour-Navier-Stokes equation has been used to 
derive a balance law for the angular momentum associated with 
each of the leaves $\Sp_t$ and a generic divergence-free vector. 
When $\Hor$ is a dynamical horizon, this law is a time differential 
form of the law obtained by Ashtekar and Krishnan \cite{AshteK03,AshteK04}. 
When $\Hor$ is a slowly evolving horizon, we recover 
the angular momentum flux law obtained by Booth and Fairhurst
\cite{BoothF04}.

\begin{acknowledgments}
It is a pleasure to thank Jos\'e Luis Jaramillo for numerous discussions
and to acknowledge the warm hospitality of the Yukawa Institute
for Theoretical Physics and the Department of Physics of Kyoto University
where this work was completed. 
\end{acknowledgments}

%%%%%%%%%%%%%%%%%%%%%%%%%%%%%%%%%%%%%%%%%%%%%%%%%%%%%%%%%%%%%%%%%%%%%%%%%%%%%%%

\appendix

\section{Link with the 3+1 description of dynamical horizons}
\label{s:AK}

\subsection{Basic relations}

When $\Hor$ is a dynamical horizon, it is a spacelike hypersurface
and Ashtekar and Krishnan \cite{AshteK03,AshteK04} have made use of
the standard 3+1 formalism to describe it, by introducing
its unit timelike future directed normal $\w{\bar n}$, its 
positive definite induced metric $\w{\bar\gamma}$ and its extrinsic
curvature tensor $\w{\bar K}$. Here we have put bars on the symbols
denoting them to stress that these objects are relative to $\Hor$
itself and not to some spacelike hypersurface $\Sigma$ intersecting
$\Hor$ in a 2-surface (as in the 3+1 perspective mentioned in
Sec.~\ref{s:perspective}). Note that our sign convention for the
extrinsic curvature is the opposite of that of Ashtekar and Krishnan 
and that we are using $\w{\bar n}$
for the unit normal denoted $\w{\hat\tau}$ by Ashtekar and Krishnan.
We also denote by $\w{\bar s}$ the unit
spacelike normal to $\Sp_t$ lying in $\Hor$. $(\w{\bar n},\w{\bar s})$
constitutes then an orthonormal frame normal to $\Sp_t$. 
$\w{\bar s}$ is denoted by $\w{\hat r}$ by Ashtekar and Krishnan,
but we privilege here notations consistent with those 
introduced in Sec.~\ref{s:normal_frame}. 
The correspondence between both sets of notations is given in
Table.~\ref{t:AK}. 
 
\begin{table}
\caption{\label{t:AK} Correspondence between our notations
and those of Ashtekar and Krishnan}
\begin{ruledtabular}
\begin{tabular}{ll}
  this work & Ashtekar and Krishnan \cite{AshteK03} \\
\hline
$\w{\bar n}$ & $\w{\hat\tau}$ \\
$\w{\bar s}$ & $\w{\hat r}$ \\
$\w{\bar\gamma}$ & $\w{q}$ \\
$\w{\bar K}$ & $-\w{K}$ \\
$\w{q}$ & $\w{\widetilde q}$ \\
$\DS$ & $\w{\widetilde D}$ \\
$\w{\Theta}^{(\w{\bar s})}$ & $\w{\widetilde K}$ \\
$\theta^{(\w{\bar s})}$ & $\widetilde K$ \\
$\w{\bar\ell}$ & $\w{\ell}$ \\
$\w{\bar k}$ & $\w{n}$ \\
$\w{\Omega}^{(\w{\bar n})}$ & $ \w{\widetilde W}$ \\
$\w{\Omega}^{(\w{\tilde\ell})}$ & $\w{\zeta}$ \\
$\frac{1}{\sqrt{2C}} \frac{dR}{dt}$ & $N_R$ \\
$\frac{1}{C} \frac{dR}{dt}\, \el
 = \frac{A}{C} \frac{dR}{dt}\, \w{\tilde\ell} $ & $\w{\xi}_{(R)}$
\end{tabular}
\end{ruledtabular}
\end{table}

The normal $\w{\bar n}$ is necessarily colinear to $\w{m}$. 
Similarly $\w{\bar s}$ is necessarily colinear to $\w{h}$.
From the norm of $\w{m}$ [Eq.~(\ref{e:norm_m})] and
$\w{h}$ [Eq.~(\ref{e:def_C})], we deduce
\be \label{e:bn_m_bs_h}
    \w{\bar n} = \frac{1}{\sqrt{2C}} \, \w{m}
    \qquad\mbox{and}\qquad
    \w{\bar s} = \frac{1}{\sqrt{2C}} \, \w{h} .    
\ee
Let us recall that $C>0$ for a dynamical horizon. 
Ashtekar and Krishnan \cite{AshteK03,AshteK04} consider the following
null normal frame (see Table~\ref{t:AK}):
\be
    \w{\bar\ell} := \w{\bar n} + \w{\bar s}
    \qquad\mbox{and}\qquad
    \w{\bar k} := \w{\bar n} - \w{\bar s} .
\ee
From Eqs.~(\ref{e:bn_m_bs_h}), (\ref{e:h_el_k}) and (\ref{e:def_m}),
we get immediately the relation between these vectors and the null vectors
$(\el,\w{k})$ associated with $\w{h}$ and introduced 
in Sec.~\ref{s:norm_frame_h}:
\be \label{e:bl_l_bk_k}
   \w{\bar\ell} = \sqrt{\frac{2}{C}} \, \el  
    \qquad\mbox{and}\qquad
    \w{\bar k} = \sqrt{2 C}\, \w{k} . 
\ee
Ashtekar and Krishnan have introduced the 1-form 
\bea
    \widetilde W_\alpha & := & 
    - {\bar K}_{\mu\nu} {\bar s}^\nu q^\mu_{\ \, \alpha}
    = ({\bar\gamma}^\rho_{\ \, \mu}{\bar\gamma}^\sigma_{\ \, \nu} 
        \nabla_\rho {\bar n}_\sigma) \, {\bar s}^\nu q^\mu_{\ \, \alpha}
        \nonumber \\
   & = & \nabla_\mu {\bar n}_\nu \, {\bar s}^\nu q^\mu_{\ \, \alpha} 
   = \Omega^{(\w{\bar n})}_\alpha,          \label{e:tW_Omega_bn}
\eea
hence $\w{\widetilde W}$ is nothing but 
the normal fundamental form $\w{\Omega}^{(\w{\bar n})}$
[cf. Eq.~(\ref{e:def_Omega_n})].

Another 1-form introduced by them is\footnote{actually they introduced it as a
vector, but we consider it here as a 1-form via the standard metric
duality}
\be
    \zeta_\alpha := {\bar s}^\mu \nabla_\mu {\bar \ell}_\nu \, q^\nu_{\ \, \alpha} .
\ee
Replacing $\w{\bar s}$ and $\w{\bar\ell}$ by their respective expressions
(\ref{e:bn_m_bs_h}) and (\ref{e:bl_l_bk_k}), and making use of 
Eq.~(\ref{e:h_el_k}), yields
\be
    \zeta_\alpha = \frac{1}{C}  \ell^\mu \nabla_\mu \ell_\nu \, q^\nu_{\ \, \alpha}
    - k^\mu \nabla_\mu \ell_\nu \, q^\nu_{\ \, \alpha} .
\ee
Thanks to the identities (\ref{e:inaff_l}) and (\ref{e:grad_k_ul}), and
to the fact that for $C>0$, Eq.~(\ref{e:const_B_CsA}) implies
$\DS\ln A = \DS\ln C$, we get
\be
    \w{\zeta} = \w{\Omega}^{(\el)} - \DS \ln C . 
\ee
Now, since $\el = A \w{\tilde\ell}$, we have from the scaling law
(\ref{e:rescale_Omega_l}),
$\w{\Omega}^{(\el)} =  \w{\Omega}^{(\w{\tilde\ell})} + \DS \ln A
= \w{\Omega}^{(\w{\tilde\ell})} + \DS \ln C$. Hence we conclude that the
quantity $\w{\zeta}$ introduced by Ashtekar and Krishnan 
\cite{AshteK03,AshteK04} is nothing but the normal fundamental form
associated with the null vector $\w{\tilde\ell}$:
\be
    \w{\zeta} = \w{\Omega}^{(\w{\tilde\ell})}  .
\ee

In Ashtekar and Krishnan analysis \cite{AshteK03,AshteK04}, a privileged
role is played by the area radius $R$, i.e. the scalar field on $\Hor$, 
which is constant on each 2-surface $\Sp_t$ and related to the area $a$ of
this surface by $a = 4\pi R^2$. An associated quantity is the lapse
$N_R$ defined as the norm of the gradient of $R$ within $\Hor$:
\be
    N_R := \sqrt{ {\bar\gamma}^{ij} \partial_i R \, \partial_j R} .
\ee 
$R$ is a function of $t$ and we may write
$\partial_i R = (dR/dt)\, \partial_i t = (dR/dt)\, {\bar s}_i / \sqrt{2C}$.
From the normalization ${\bar\gamma}^{ij} {\bar s}_i {\bar s}_j = 1$
and the positivity of $dR/dt$ (area increase law \cite{AshteK03}),
we then obtain
\be \label{e:N_R_AK}
    N_R = \frac{1}{\sqrt{2C}} \frac{dR}{dt} . 
\ee
The null evolution vector considered by Ashtekar and Krishnan
is $\w{\xi}_{(R)} = N_R \w{\bar\ell}$. From Eqs.~(\ref{e:bl_l_bk_k}),
(\ref{e:N_R_AK}), and (\ref{e:elk_tilde_elk}), we can re-express it as
\be
   \w{\xi}_{(R)} = \frac{1}{C} \frac{dR}{dt}\, \el
   = \frac{A}{C} \frac{dR}{dt}\, \w{\tilde\ell} . 
\ee
Notice that thanks to the property (\ref{e:const_B_CsA}), 
the coefficient $A/C\, dR/dt$ in front of $\w{\tilde\ell}$
is constant on each 2-surface $\Sp_t$. 

\subsection{Angular momentum}

Ashtekar and Krishnan \cite{AshteK03,AshteK04} define the generalized
angular momentum associated with a section $\Sp_t$ and a vector field
$\w{\varphi}$ tangent to $\Sp_t$ by
\be
    {\bar J}(\w{\varphi}) :=  \frac{1}{8\pi} \oint_{\Sp_t} 
    \w{\bar K}(\w{\varphi},\w{\bar s}) \, \volS . 
\ee
From Eq.~(\ref{e:tW_Omega_bn}) and $\vec{\w{q}}(\w{\varphi}) = \w{\varphi}$, 
we deduce immediately that 
\be
    {\bar J}(\w{\varphi}) =  - \frac{1}{8\pi} \oint_{\Sp_t} 
    \langle \w{\Omega}^{(\w{\bar n})}, \w{\varphi} \rangle  \, \volS . 
\ee
If we suppose now that $\w{\varphi}$ is divergence-free with respect
to the connection in $\Sp_t$: $\DS\cdot\w{\varphi}=0$, 
then, by means of the transformation laws of normal fundamental forms, Eqs.~
(\ref{e:Omega_el_n}) and (\ref{e:rescale_Omega_l}), 
it is easy to see that ${\bar J}(\w{\varphi})$ coincides with 
the generalized angular momentum $J(\w{\varphi})$ 
as defined by Eq.~(\ref{e:def_genJ}):
\be
    \DS\cdot\w{\varphi}=0 \ \Longrightarrow \ {\bar J}(\w{\varphi})
    = J(\w{\varphi}) . 
\ee 

Regarding the angular momentum flux law, Ashtekar and Krishnan \cite{AshteK03}
have derived an integrated version of it from the momentum constraint
equation relative to the hypersurface $\Hor$. It writes
\begin{widetext}
\be 
     J(\w{\varphi},t_2) - J(\w{\varphi},t_1)
    = -\int_{\Delta\Hor} \w{T}(\w{\bar n},\w{\varphi}) \; \volH
        - \frac{1}{16\pi} \int_{\Delta\Hor}  
       \left[ ({\bar K}{\bar \gamma}^{ij}- {\bar K}^{ij}) \Liec{\varphi}
       {\bar\gamma}_{ij} \right]\, \volH , \label{e:AK_flux_law}
\ee
where $\Delta\Hor$ is a portion of $\Hor$ delimited by two surfaces,
$\Sp_{t_1}$ and $\Sp_{t_2}$ say, and $\volH$ is the volume 3-form on 
$\Hor$ associated with the metric $\w{\bar\gamma}$. 
Note that we have restored the explicit dependence of $J(\w{\varphi})$ on
$\Sp_t$ by writing $J(\w{\varphi},t)$.
Note also that Eq.~(\ref{e:AK_flux_law}) holds for any spacelike hypersurface
$\Hor$, not necessarily a dynamical horizon. 
Let us express the integrand in the second integral
in the right-hand side of Eq.~(\ref{e:AK_flux_law}) in terms of fields
defined on the 2-surfaces $\Sp_t$. First of all, performing an orthogonal
2+1 decomposition of $\w{\bar K}$ with respect to $\Sp_t$ yields
\bea
    {\bar K}^{ij} & =&  - \Theta^{(\w{\bar n})ij} 
    - \Omega^{(\w{\bar n})i} {\bar s}^j  - \Omega^{(\w{\bar n})j} {\bar s}^i
    + ({\bar K}_{kl} {\bar s}^k {\bar s}^l) \, {\bar s}^i {\bar s}^j  
            \label{e:bKij_2p1} \\
    {\bar K} & = & - \theta^{(\w{\bar n})}
        +  {\bar K}_{kl} {\bar s}^k {\bar s}^l  . \label{e:bK_2p1}
\eea
Besides, $\Liec{\varphi} {\bar\gamma}_{ij} = {\bar D}_i \varphi_j
    + {\bar D}_j \varphi_i$, where ${\bar D}_i$ is the covariant derivative
    associated with the 3-metric $\w{\bar\gamma}$ on $\Hor$ and 
    $\varphi_i := \bar\gamma_{ij} \varphi^j$ ,
with
\be
    {\bar D}_i \varphi_j = \DSc_i \varphi_j - \Theta^{(\w{\bar s})}_{ik} 
        \varphi^k {\bar s}_j
        + {\bar s}_i {\bar s}^k {\bar D}_k \varphi_j .
\ee
Using the property $\Lie{h}\w{\varphi}=0$ [Eq.~(\ref{e:varphi_htransp})]
and the relation (\ref{e:bn_m_bs_h}) between $\w{h}$ and $\w{\bar s}$,
we can rewrite the above expression as
\be \label{e:Dphi_2p1}
    {\bar D}_i \varphi_j = \DSc_i \varphi_j + \left( {\bar s}_i
        \Theta^{(\w{\bar s})}_{jk} 
        - {\bar s}_j \Theta^{(\w{\bar s})}_{ik} \right) \varphi^k 
        +\frac{1}{2} \varphi^k \DSc_k\ln C \,  {\bar s}_i {\bar s}_j  .
\ee
From Eqs.~(\ref{e:bKij_2p1}), (\ref{e:bK_2p1}) and (\ref{e:Dphi_2p1}) and
the divergence-free property of $\w{\varphi}$ [Eq.~(\ref{e:div_varphi_zero})], 
we get 
\be
    ({\bar K}{\bar \gamma}^{ij}- {\bar K}^{ij}) \Liec{\varphi}
       {\bar\gamma}_{ij} =  2 \sigma^{(\w{\bar n})ab} \DSc_a \varphi_b
       - \theta^{(\w{\bar n})} \varphi^a \DSc_a \ln C = \
\stackrel{\twoheadrightarrow}{\w{\sigma}}\!\!{}^{(\w{\bar n})}\!:\Lie{\varphi}\w{q}
 -  \theta^{(\w{\bar n})}  \w{\varphi} \cdot \DS \ln C .     
\ee
Now from Eq.~(\ref{e:bn_m_bs_h}), 
$\w{\sigma}^{(\w{\bar n})} = \w{\sigma}^{(\w{m})} / \sqrt{2C}$
and $\theta^{(\w{\bar n})} = \theta^{(\w{m})} / \sqrt{2C}$, so that we
can write Ashtekar and Krishnan's integrated flux law 
(\ref{e:AK_flux_law}) as
\be 
     J(\w{\varphi},t_2) - J(\w{\varphi},t_1)
    = -\int_{\Delta\Hor} \frac{1}{\sqrt{2C}}\w{T}(\w{m},\w{\varphi}) \; \volH
         - \frac{1}{16\pi} \int_{\Delta\Hor}  
     \frac{1}{\sqrt{2C}}
     \left[ 
 \stackrel{\twoheadrightarrow}{\w{\sigma}}\!\!{}^{(\w{m})}\!:\Lie{\varphi}\w{q}
     - \theta^{(\w{m})} \w{\varphi} \cdot \DS \ln C  \right]
                                \; \volH . \label{e:AK_flux_law2}
\ee
On the other side, if we integrate in time our flux law 
(\ref{e:Jflux_non_null}), which, as Eq.~(\ref{e:AK_flux_law2}), 
is valid for any spacelike hypersurface $\Hor$, we get 
\be 
     J(\w{\varphi},t_2) - J(\w{\varphi},t_1)
    = - \int_{\Delta\Hor} \w{T}(\w{m},\w{\varphi}) \; 
    {}^{\scriptscriptstyle\Hor}\!{\bm{\mathrm{d}}} t \wedge \volS
            - \frac{1}{16\pi} \int_{\Delta\Hor}  
    \left[
 \stackrel{\twoheadrightarrow}{\w{\sigma}}\!\!{}^{(\w{m})}\!:\Lie{\varphi}\w{q}
    - 2 \theta^{(\el)} \w{\varphi} \cdot \DS \ln C  \right]   
       \; {}^{\scriptscriptstyle\Hor}\!{\bm{\mathrm{d}}} t \wedge \volS ,
                        \label{e:flux_integr}
\ee
\end{widetext}
where ${}^{\scriptscriptstyle\Hor}\!{\bm{\mathrm{d}}} t$ denotes the gradient
1-form of the scalar field $t$ within the manifold $\Hor$. 
Besides, from the basic properties $\Lie{h}{t}=1$, $\w{h}$ orthogonal
to $\Sp_t$ and $\w{h}\cdot\w{h}=2C$, we deduce easily that 
$\underline{\w{h}} := \w{\bar\gamma}\cdot\w{h} = 
2C \, {}^{\scriptscriptstyle\Hor}\!{\bm{\mathrm{d}}} t$, 
hence 
$\underline{\w{\bar s}} := \w{\bar\gamma}\cdot\w{\bar s}
= \underline{\w{h}} / \sqrt{2C}
= \sqrt{2C} \, {}^{\scriptscriptstyle\Hor}\!{\bm{\mathrm{d}}} t$.
Now, since $\w{\bar s}$ is the unit vector orthogonal to $\Sp_t$,
$\volH = \underline{\w{\bar s}} \wedge \volS$, so that we have
\be
    \volH = \sqrt{2C} \; {}^{\scriptscriptstyle\Hor}\!{\bm{\mathrm{d}}} t
       \wedge \volS . 
\ee
This relation shows the equivalence of Eqs.~(\ref{e:AK_flux_law2})
and (\ref{e:flux_integr}), except at first glance for the $\theta^{(\w{m})}$
term in Eq.~(\ref{e:AK_flux_law2}) which is replaced by $2\theta^{(\el)}$
in Eq.~(\ref{e:flux_integr}). However, 
$\theta^{(\w{m})} = 2 \theta^{(\el)} - \theta^{(\w{h})}$
[cf. Eqs.~(\ref{e:h_el_k}) and (\ref{e:def_m})]
and the integral over $\Sp_t$ of 
$\theta^{(\w{h})} \w{\varphi} \cdot \DS \ln C $ vanishes since
the vector field $\theta^{(\w{h})} \w{\varphi}$ is divergence-free
[Eq.~(\ref{e:th_vp_divfree})]. 
This proves that Eq.~(\ref{e:flux_integr}) is identical to 
Eq.~(\ref{e:AK_flux_law2}), i.e. that for a spacelike hypersurface, and in 
particular for a dynamical horizon, the integrated version of our 
angular momentum flux law
(\ref{e:Jflux_non_null}) results in Ashtekar and Krishnan \cite{AshteK03}
angular momentum balance equation.

\end{document}